\documentclass[11pt,psamsfonts]{amsart}
\usepackage{amsmath}
\usepackage{amsthm}
\usepackage{amssymb}
\usepackage{amscd}
\usepackage{amsfonts}
\usepackage{amsbsy}
\usepackage{booktabs}
\usepackage{calc}
\usepackage{caption}
\usepackage{color}
\usepackage{graphicx}
\usepackage{lineno}
\usepackage{lscape}
\usepackage[misc]{ifsym}
\usepackage{multirow}
\usepackage{setspace}
\usepackage{subfigure}

\usepackage{geometry}
\begin{document}
%\linenumbers
\title[]
{Revisiting the distributions of Jupiter's irregular moons: I. physical characteristics}
 \author{Fabao~Gao$^{1,2}$, Xia~Liu$^{1}$}
 \address{${}^1$School of Mathematical Science, Yangzhou University, Yangzhou 225002, China}
  \address{${}^2$Departament de Matem$\grave{\text{a}}$tiques, Universitat Aut$\grave{\text{o}}$noma de Barcelona, Bellaterra 08193, Barcelona, Catalonia, Spain}
  \address{\textnormal{E-mail: gaofabao@sina.com ( \Letter\ Fabao Gao, ORCID 0000-0003-2933-1017), liuxiayzu@sina.com (Xia Liu)}}


\begin{abstract}
As the identified number of Jupiter's moons has skyrocketed to 79, some of them have been regrouped. In this work, we continue to identify the potential distributions of the physical characteristics of Jupiter's irregular moons. By using nonparametric Kolmogorov-Smirnov tests, we verified more than 20 commonly used distributions and found that surprisingly, almost all the physical characteristics (i.e., the equatorial radius, equatorial circumference, circumference, volume, mass, surface gravity and escape velocity) of the moons in the Ananke and Carme groups follow log-logistic distributions. Additionally, more than half of the physical characteristics of the moons in the Pasiphae group are theoretically subject to this type of distribution. The discovery of an increasing number of Jupiter's irregular moons combined with strict analytical derivations, it is increasingly clear and possible to anticipate that the physical characteristics of most irregular moons follow log-logistic distributions.
\end{abstract}
\maketitle

\section{Introduction}

As the largest planet in the solar system, Jupiter is known to have at least 79 moons (see [1] for more details); this structure is similar to that of the solar system. The study of Jupiter and its moons may help explain the origin and evolution of Jupiter, its moons and even the solar system. In recent years, many researchers have shown great interest in the distribution and origin of Jupiter's natural satellites (see [2]-[9] and the references therein). 

Generally, Jupiter's natural satellites can be grouped into regular moons and irregular moons according to their orbital inclination and direction of rotation around the planet. The regular moons have nearly circular prograde orbits and low inclination, which suggests that their orbits are close to Jupiter's equator. In contrast, the orbits of the irregular moons have relatively high eccentricity and inclination, and they are far from Jupiter and often follow retrograde orbits. These irregular moons are believed to have been at least partially formed by the collision of asteroids captured by Jupiter's gravitational field [3]-[5] or by the complex gravitational interaction of the several giant planets [6]-[7]. Ronnet et al. [8] hypothesized that there is a planetesimal reservoir at the outer edge of Jupiter's gap (a notch or dip in the density distribution of the gas surface), where the captured solids are trapped and Jupiter's moons are gradually formed. The capture of the solids may have been caused by the energy loss of a planetesimal inside Jupiter’s Hill sphere. In addition to the captured irregular satellites, there is the possibility of irregular satellite formation/collision. Several compact clusters of orbits around Jupiter, supported by color similarities, have been indicated to have a common origin [10].

In 1766, Titius proposed a simple geometric rule based on the orbital radii of the planets in the solar system, which was then summarized by Bode of the Berlin Observatory into an empirical formula called the Titius-Bode law. This formula correctly predicted the orbits of Uranus and Ceres in the asteroid belt [14]-[17] and was later extended by other researchers [11]-[13]. However, due to the lack of strict theoretical explanations, many astronomers now believe that this law is just a coincidence. It has even been rejected by some researchers, although similar issues in other planetary systems remain unresolved [18].

Although we also agree that the Titius-Bode law maybe a pure assumption without a strictly theoretical basis, it can be adapted to study the distributions of some regular moons. Over the past year, with the discovery of a large number of Jupiter's moons, we are motivated to investigate the distributions of the physical characteristics of Jupiter's irregular moons. Of course, we do not present a hypothesis but rather strictly examine whether there are some statistical laws governing the irregular moons. Previously, there were only 45 irregular members in the three major groups of Jupiter's moons (see [19] for details): 15 moons in the Carme group, 11 moons in the Ananke group and 19 moons in the Pasiphae group. Gao et al. [2] proposed that the $t$ location-scale distribution and Weibull distribution are the main distributions of the four physical characteristics of the equatorial radius, mass, surface gravity and escape velocity. Moreover, they believed that if future observations allow for an increase in the number of Jupiter's moons, the distributions may change slightly but would not change significantly for a long time. The total number of irregular moons has increased from the previous 45 to the current 54 in Jupiter's three major groups, of which 14 moons have been updated: 9 have been newly discovered and 5 have been regrouped. More precisely, the Carme group accepted 5 newly discovered moons (S/2003 J19, S/2011 J1, S/2017 J2, S/2017 J5 and S/2017 J8), for a total of 20 moons. In addition to the unique new member S/2017 J16 in the Pasiphae group, 5 senior members (S/2001 J9, S/2001 J10, S/2003 J6, S/2003 J18, and S/2016 J1) were transferred out of this group and into the Ananke group, so the total number of satellites in the Pasiphae group was reduced to 15 moons. In addition to the 5 newly accepted satellites in the Ananke group, 3 newly discovered members (S/2017 J3, S/2017 J7 and S/2017 J9) joined the Ananke group, vaulting the total number of moons in this group to 19. 

In this paper, based on the above three updated major groups of Jupiter's moons and 21 commonly used distribution functions (see Appendix A), we investigate the distribution rules with respect to the seven physical characteristics of irregular moons by using one-sample nonparametric Kolmogorov-Smirnov (K-S) tests (see [20] for details) and the maximum likelihood estimation method. These seven physical characteristics are the equatorial radius, equatorial circumference, volume, surface area, surface gravity, mass and escape velocity. In addition, it is noted that the seven physical characteristics are not completely independent, which will help us verify the rationality of the results obtained by statistical inference through strict analytical derivation. 

\section{Principle of statistical inference}
A nonparametric test is a method to infer the type of population distribution by using sample data when the population variance is unknown or poorly understood. Because this method does not involve the parameters of a population distribution in the process of inference, it is called a nonparametric test. This method can be used to infer whether the population from which the sample comes follows a certain theoretical distribution.

If $F_n (x)=i/n\ (i = 1, 2, \cdots, n)$ is a frequency distribution function of observations from $n$ random samples, then it represents the number of all observations less than or equal to the value of $x$. $F (x)$ denotes the theoretical cumulative distribution function (CDF) of a population; i.e., the value of $F (x)$ represents the proportion of expected results that are less than or equal to the value of $x$. The aforementioned K-S test is a nonparametric test method that compares the frequency distribution $F_n (x)$ with the theoretical distribution $F (x)$ or the distribution of two observations. The difference between $F_n (x)$ and $F (x)$ is defined as follows:
\begin{equation}
\begin{array}{rl}
D_n = \sup_x|F_n (x)-F (x) |.
\end{array}
\end{equation}
If $F_n (x)$ and $F (x)$ are close enough for each value of $x$, i.e., $D_n\to0$ when $n\to+\infty$, then these two functions have a high degree of fitting, and it is reasonable to believe that the sample data come from the population following the theoretical distribution $F (x)$. The K-S test focuses on the largest deviation in equation (1), and an evaluation can be made with the help of constructed statistics. To this end, two opposite hypotheses are proposed---namely, the null hypothesis $H_0$ and the alternative hypothesis $H_1$. $H_0$ is the hypothesis that the sample comes from a specific distribution. It is an event that is likely to occur in one test. The alternative hypothesis $H_1$ is usually assigned to an event that is highly unlikely to occur in one test. The detailed procedures of the K-S test are listed as follows:

(I). Establish the hypotheses:
$H_0: F_n (x) = F (x),\ H_1: F_n (x) \neq F (x)$.

(II). Calculate the statistics: $D_n$.

(III). Determine the critical value: The critical value $D_{n,\alpha}$ can be obtained according to the given significance level $\alpha$ (usually set to 0.05 or 0.01) and the sample size $n$.

(IV). Make the judgment: $H_0$ cannot be rejected at the $\alpha$ level when $D_n <D_{n,\alpha}$. Otherwise, $H_0$ should be rejected.

It is noted from step (III) that the level $\alpha$ must be specified first before one can continue the procedure. However, when the value of $\alpha$ decreases, the rejection domain of the test correspondingly decreases, causing the observation value that initially fell into the rejection domain to eventually fall into the acceptance domain, so this case sometimes causes problems in practical applications. To avoid the inconvenience caused by predetermining $\alpha$, the $p$-value is introduced, and then one can easily draw more intuitive conclusions about the test by comparing the $p$-value and $\alpha$. Here, the $p$-value indicates the minimum significance level at which the null hypothesis can be rejected according to the sample values of the test statistics. The smaller the $p$-value, the stronger the evidence against $H_0$. Thus, if $p \leq \alpha$, $H_0$ is rejected at the $\alpha$ significance level. Otherwise, we fail to reject $H_0$ if $p> \alpha$. Furthermore, if the $p$-values ​​corresponding to several distributions are all greater than $\alpha$ and the differences are large, the distribution with the largest $p$-value is selected as the best-fit inferred distribution. If all $p$-values ​​are greater than $\alpha$ and close to each other, the confidence interval is also considered, and the corresponding distribution is selected as the best-fit distribution.

\section{Distribution inference based on different groups of Jupiter's moons}

\subsection{Ananke group}
Based on one-sample K-S tests and maximum likelihood estimation, the results of statistical inference on the aforementioned seven physical characteristics of irregular moons can be found in Tables A1-A4 in Appendix A. The best-fit distributions of the physical characteristics are summarized in Table 1. All the physical characteristics follow a log-logistic distribution, and their $p$-values are approximately 0.8, except that the p-value for the surface gravity is less than 0.5 but much greater than 0.05. Therefore, it is clear that the log-logistic distribution is the best-fit distribution to describe the physical characteristics in the Ananke group. The probability distribution function (PDF) of the log-logistic distribution has the following form:
\begin{equation}
\begin{aligned}
f(x)=\frac{1}{\sigma x }\frac{e^{\frac{\ln\left (x\right )-\mu }{\sigma }}}{\left [ 1+e^{\frac{\ln\left (x\right) -\mu }{\sigma }} \right ]^{2}},\label{eq23}
\end{aligned}
\end{equation}
where $\mu$ is the mean of the logarithmic values and $\sigma$ is the scale parameter of the logarithmic values. Here, it should be noted that this function is not unique, and physical characteristics can also be fitted by other functions. It is believed that with the advancement of astronomical observation technology and the development of statistics, this function may be replaced in the future.

As seen in Table 1, the $p$-values (only four decimal places are retained) of the equatorial radius ($R$), equatorial circumference ($C$), volume ($V$) and surface area ($S$) are approximately 0.83. A reasonable explanation is that there is a linear relationship $C=2\pi R$ between the equatorial radius and the equatorial circumference. However, it may be a coincidence that the $p$-values corresponding to the nonlinear relationships $V = 4 \pi R^3 / 3$ and $S = 4 \pi R^2$ are also close to each other because their differences emerge in the Carme and Pasiphae groups.

%Table 1
\begin{table*}[!htb]
\newcommand{\tabincell}[2]{\begin{tabular}{@{}#1@{}}#2\end{tabular}}
\centering

\caption{\label{opt}Inference of the distribution of each physical characteristic for the moons in the Ananke group}
\footnotesize
\rm
\setlength{\tabcolsep}{3.6mm}{
\centering
\begin{tabular}{@{}*{7}{l}}
 \toprule
\textbf{Characteristic}&\textbf{Best-fit Distribution}&\textbf{Parameters}&\textbf{Confidence Intervals}&\textbf{$p$-value}\\
 \toprule

Equatorial Radius ($km$) & Loglogistic& \tabincell{l}{$\mu$=0.582917\\$\sigma$=0.364002 }& \tabincell{l}{$\mu$$\in$[0.23278, 0.933054]\\$\sigma$$\in$[0.225066, 0.588705]} & 0.8343  \\
\hline
Equatorial Circumference ($km$) & Loglogistic & \tabincell{l}{$\mu$=2.42182\\$\sigma$=0.363511 }& \tabincell{l}{$\mu$$\in$[2.07218, 2.77145]\\$\sigma$$\in$[0.224753, 0.587933] }& 0.8327  \\
\hline
Volume ($km^{3}$)&Loglogistic & \tabincell{l}{$\mu$=3.17586\\$\sigma$=1.10596 }& \tabincell{l}{$\mu$$\in$[2.11064, 4.24108]\\$\sigma$$\in$[0.684312, 1.7874] }& 0.8339  \\
\hline
Surface Area ($km^{2}$) & Loglogistic & \tabincell{l}{$\mu$=3.69694\\$\sigma$=0.727939 }& \tabincell{l}{$\mu$$\in$[2.99674, 4.39715]\\$\sigma$$\in$[0.45009, 1.17731] }& 0.8342  \\
\hline
Surface Gravity ($m/s^{2}$)& Loglogistic& \tabincell{l}{$\mu$=-6.41229\\$\sigma$=0.337578} & \tabincell{l}{$\mu$$\in$[-6.74044, -6.08413]\\$\sigma$$\in$[0.210095, 0.542414] }& 0.4831  \\
\hline
Mass ($kg$) & Loglogistic & \tabincell{l}{$\mu$=31.8851\\$\sigma$=1.00964 }& \tabincell{l}{$\mu$$\in$[30.921, 32.8493]\\$\sigma$$\in$[0.621833, 1.63929]} & 0.7825  \\
\hline
Escape Velocity ($km/h$) & Loglogistic& \tabincell{l}{$\mu$=2.10984\\$\sigma$=0.329425 }& \tabincell{l}{$\mu$$\in$[1.79575, 2.42393]\\$\sigma$$\in$[0.202763, 0.535212] }& 0.7718  \\

\toprule
\end{tabular}}
 \label{tab3-1} 
\end{table*}

\subsection{Carme group}

For the Carme group (see Table 2 and Tables A5-A8 in Appendix A for more details), the $p$-values corresponding to the escape velocity in the $t$ location-scale distribution and the log-logistic distribution are close to each other, approximately 0.5996 and 0.5222, respectively. However, the confidence interval corresponding to a $t$ location-scale distribution is more dispersed than that of a log-logistic distribution; thus, the best-fit distribution of the escape velocity may be a $t$ location-scale distribution or log-logistic distribution. The remaining six physical characteristics all follow log-logistic distributions. Similar to the results for the Ananke group, the $p$-values of the first two physical characteristics (i.e., the equatorial radius and equatorial circumference) in the Carme group are approximately 0.66 because they are not independent and have a linear relationship.
%Table 2
\begin{table*}[!htb]
\newcommand{\tabincell}[2]{\begin{tabular}{@{}#1@{}}#2\end{tabular}}
\centering

\caption{\label{opt}Inference of the distribution of each physical characteristic for the moons in the Carme group}
\footnotesize
\rm
\setlength{\tabcolsep}{3.6mm}{
\centering
\begin{tabular}{@{}*{7}{l}}
 \toprule
\textbf{Characteristic}&\textbf{Best-fit Distribution}&\textbf{Parameters}&\textbf{Confidence Intervals}&\textbf{$p$-value}\\
 \toprule
Equatorial Radius ($km$)& Loglogistic &  \tabincell{l}{$\mu$=0.500116\\$\sigma$=0.345753} &  \tabincell{l}{$\mu$$\in$[0.191027, 0.809204]\\$\sigma$$\in$[0.21508, 0.555817]} & 0.6654  \\
    \hline
Equatorial Circumference ($km$)& Loglogistic &  \tabincell{l}{$\mu$=2.33795\\$\sigma$=0.344919} &  \tabincell{l}{$\mu$$\in$[2.0297, 2.6462]\\$\sigma$$\in$[0.214533, 0.55455]} & 0.6612  \\
    \hline
Volume ($km^{3}$) & Loglogistic&  \tabincell{l}{$\mu$=2.9226\\$\sigma$=1.04959} &  \tabincell{l}{$\mu$$\in$[1.98296, 3.86223]\\$\sigma$$\in$[0.653374, 1.68606] }& 0.6817  \\
    \hline
Surface Area ($km^{2}$)  &Loglogistic &  \tabincell{l}{$\mu$=3.53128\\$\sigma$=0.691438 }&  \tabincell{l}{$\mu$$\in$[2.91317, 4.14939]\\$\sigma$$\in$[0.430115, 1.11153] }& 0.6651  \\
    \hline
Surface Gravity ($m/s^{2}$)& Loglogistic&  \tabincell{l}{$\mu$=-6.69388\\$\sigma$=0.319434 }&  \tabincell{l}{$\mu$$\in$[-6.97911, -6.40864]\\$\sigma$$\in$[0.197127, 0.517626] }& 0.0584  \\
   \hline
Mass ($kg$) & Loglogistic&  \tabincell{l}{$\mu$=31.6362\\$\sigma$=0.956765} &  \tabincell{l}{$\mu$$\in$[30.7891, 32.4832]\\$\sigma$$\in$[0.59232, 1.54545]} & 0.5776  \\
    \hline
Escape Velocity ($km/h$)&\tabincell{l}{$t$ location-scale\\(Loglogistic)} &  \tabincell{l}{$\mu$=7.08058\\$\sigma$=1.51796\\$\nu$=0.993412\\($\mu$=2.01252\\$\sigma$=0.30989) }&  \tabincell{l}{$\mu$$\in$[5.84963, 8.31154]\\$\sigma$$\in$[0.660915, 3.4864]\\$\nu$$\in$[0.403482, 2.44588]\\($\mu$$\in$[1.73925, 2.2858]\\$\sigma$$\in$[0.191438, 0.501634])} &  \tabincell{l}{0.5996\\(0.5222)}\\
\toprule
\end{tabular}}
 \label{tab3-2} 
\end{table*}

\subsection{Pasiphae group}

In the Pasiphae group (see Table 3 and Tables A9-A12 in Appendix A for more details), all the best-fit distributions of the equatorial radius, equatorial circumference, surface gravity and escape velocity are still log-logistic distributions. The volume follows a generalized Pareto distribution, which has been widely used in many fields, such as extreme value analysis, insurance loss fitting and financial risk management [21]-[22]. This distribution has three parameters: $\sigma$ is the scale parameter, $k$ is the shape parameter, and $\theta$ is the position parameter. The surface area and mass follow inverse Gaussian distributions with two parameters: $\mu$ represents the mean, and  $\lambda $ is a shape parameter.
However, $ V $ and $ S $ in this group obviously have larger and different $p$-values. Although the first two physical characteristics ($R$ and $C$) in the Pasiphae group are the same as those in the Ananke and Carme groups, both characteristics follow log-logistic distributions, and the $p$-values in the respective groups are approximately equal due to their linear relationships.
%Table 3
\begin{table*}[!htb]
\newcommand{\tabincell}[2]{\begin{tabular}{@{}#1@{}}#2\end{tabular}}
\centering
\caption{\label{opt}Inference of the distribution of each physical characteristic for the moons in the Pasiphae group}
\footnotesize
\rm
\setlength{\tabcolsep}{3.6mm}{
\centering
\begin{tabular}{@{}*{7}{l}}
 \toprule
\textbf{Characteristic}&\textbf{Best-fit Distribution}&\textbf{Parameters}&\textbf{Confidence Intervals}&\textbf{$p$-value}\\
 \toprule
Equatorial Radius ($km$)& Loglogistic & \tabincell{l}{ $\mu$=0.802683\\$\sigma$= 0.59951} & \tabincell{l}{ $\mu$$\in$[0.193211, 1.41215]\\$\sigma$$\in$[0.358041, 1.00383] }& 0.6575  \\
    \hline
Equatorial Circumference ($km$) & Loglogistic & \tabincell{l}{ $\mu$=2.64143\\$\sigma$=0.599202 }&  \tabincell{l}{$\mu$$\in$[2.03226, 3.2506]\\$\sigma$$\in$[0.357861, 1.0033] }& 0.6553  \\
    \hline
Volume ($km^{3}$)& Generalized Pareto& \tabincell{l}{ k=2.99649\\$\sigma$=17.5719\\$\theta$=0 }& \tabincell{l}{ k$\in$[0.797871, 5.19512]\\$\sigma$$\in$[4.20131, 73.4945]\\$\theta$=0 }& 0.8840  \\
     \hline
Surface Area ($km^{2}$)  &Inverse Gaussian & \tabincell{l}{ $\mu$=1487.72\\$\lambda $=30.7839 }&  \tabincell{l}{$\mu$$\in$[-4624.14, 7599.58]\\$\lambda $$\in$[5.05682, 56.511] }& 0.7373  \\
    \hline
Surface Gravity ($m/s^{2}$)& Loglogistic & \tabincell{l}{ $\mu$=-6.2884\\$\sigma$=0.555284 }& \tabincell{l}{ $\mu$$\in$[-6.85117, -5.72564]\\$\sigma$$\in$[0.331098, 0.931266]} & 0.4431  \\
   \hline
Mass ($kg$)& Inverse Gaussian &  \tabincell{l}{$\mu$$=3.41874\ast 10^{16}$\\$\lambda $=$4.03647\ast 10^{13}$ }&  \tabincell{l}{$\mu$$\in$[$-5.53893\ast 10^{17}$, \\$6.22268\ast 10^{7}$]\\$\lambda $$\in$[$6.63069\ast 10^{12}$, \\$7.40988\ast 10^{17}$] }& 0.8481  \\
  \hline
Escape Velocity ($km/h$)& Loglogistic &  \tabincell{l}{$\mu$=2.32956\\$\sigma$=0.570971 }&  \tabincell{l}{$\mu$$\in$[1.75079, 2.90834]\\$\sigma$$\in$[0.339631, 0.959888] }& 0.5832  \\
\toprule
\end{tabular}}
 \label{tab3-3} 
\end{table*}

To facilitate comparisons, in Table 4, we summarize the best-fit distributions of the seven physical characteristics for the three groups. In the Ananke and Carme groups, almost all these characteristics follow log-logistic distributions, and the physical characteristics $ R $, $ C $, and $ S $ and the escape velocity also follow this distribution in the Pasiphae group. However, we note that although the physical characteristics $S$ and mass follow an inverse Gaussian distribution, $V$ follows a generalized Pareto distribution in the Pasiphae group. It can also be found from Tables A10 and A11 in Appendix A that these three physical characteristics have large corresponding $p$-values of 0.6573, 0.5969 and 0.6707 for the log-logistic distribution, respectively. It is believed that with the discovery of new moons in this group in the future, the log-logistic distribution will also be one of the best-fit distributions.

%Table 4
\begin{table}[!htb]
\newcommand{\tabincell}[2]{\begin{tabular}{@{}#1@{}}#2\end{tabular}}
\centering
\caption{\label{opt3}Best-fit distribution inference summary}
\footnotesize
\rm
\centering
\setlength{\tabcolsep}{8mm}{
\begin{tabular}{@{}*{7}{l}}
 \toprule
\textbf{Characteristic}&\textbf{Ananke group}&\textbf{Carme group}&\textbf{Pasiphae group}\\
 \toprule
Equatorial Radius ($km$)&Loglogistic&Loglogistic&Loglogistic\\
Equatorial Circumference ($km$) &Loglogistic&Loglogistic&Loglogistic\\
Volume ($km^{3}$) &Loglogistic&Loglogistic & Generalized Pareto\\
Surface Area ($km^{2}$)  &Loglogistic&Loglogistic&Inverse Gaussian \\
Surface Gravity ($m/s^{2}$) &Loglogistic&Loglogistic&Loglogistic\\
Mass ($kg$) &Loglogistic&Loglogistic&Inverse Gaussian \\
Escape Velocity ($km/h$)&Loglogistic& \tabincell{l}{$t$ location-scale\\(Loglogistic)}&Loglogistic\\
\toprule
\end{tabular}}
 \label{tab3-4} 
\end{table}

\section{Comparison of the best-fit distributions of physical characteristics with [2]}
Since five members of the three major satellite groups have been regrouped and nine new members have joined, the best-fit distribution of the physical characteristics obtained in this paper is compared with that in [2] in this section. Note that the data in [2] use a different notation than this paper; that is, the data corresponding to the surface gravity and mass are multiplied by $10^2$ and divided by $10^{13}$, respectively. To be consistent with [2], we also work with the data accordingly. Due to the large difference in $p$-value between the two papers, we also retain only four decimal places for the $p$-values in Table 5, in this case corresponding to $0.0000$. In addition, only nine continuous distributions were selected in [2] to test the equatorial radius, surface gravity, mass and escape velocity of the four physical characteristics, without loss of generality. The 21 conventional distributions mentioned in Appendix A are still used here.

%Table 5
%\begin{landscape}
\begin{table}[tbp]
\caption{\label{opt4}Best-fit distributions of the same physical characteristics in the present paper and reference [2]}
\centering
\scalebox{0.84}{
\begin{tabular}{lllllll}
 \toprule
\multicolumn{1}{c}{\textbf{}} & \multicolumn{3}{l}{\textbf{Present paper}} & \multicolumn{3}{c}{\textbf{Reference [2]}} \\
 \toprule
\textbf{Characteristics in Ananke}&\textbf{Best-fit}&\textbf{Parameters}&\textbf{$p$-value}&\textbf{Best-fit}&\textbf{Parameters}&\textbf{$p$-value}\\
 \toprule
Equatorial Radius ($km$) & Loglogistic 
&  \begin{tabular}[c]{@{}l@{}}$\mu$=0.582917\\$\sigma$=0.364002\end{tabular}
&  \begin{tabular}[c]{@{}l@{}}0.8343\end{tabular} 
& $t$ location-scale 
&  \begin{tabular}[c]{@{}l@{}}$\mu$= 1.8072\\$\sigma$= 0.66029\\$\nu$ = 1.19615 \end{tabular}
&  \begin{tabular}[c]{@{}l@{}}0.7231\end{tabular} \\
Surface Gravity ($10^{-2}m/s^{2}$)
& Loglogistic 
&  \begin{tabular}[c]{@{}l@{}}$\mu$=-1.80711\\$\sigma$=0.337578\end{tabular}
&  \begin{tabular}[c]{@{}l@{}}0.4831\end{tabular} 
& $t$ location-scale 
&  \begin{tabular}[c]{@{}l@{}} $\mu$= 0.167044\\$\sigma$= 0.063608\\$\nu$ = 1.40815\end{tabular}
&  \begin{tabular}[c]{@{}l@{}}0.4210\end{tabular}\\
Mass ($10^{13}kg$) & Loglogistic 
&  \begin{tabular}[c]{@{}l@{}}$\mu$=1.95154\\$\sigma$=1.00964\end{tabular}
&  \begin{tabular}[c]{@{}l@{}}0.7825\end{tabular}
& Weibull 
&  \begin{tabular}[c]{@{}l@{}}$A$ = 38.3573\\$B$ = 0.368827\end{tabular}
&  \begin{tabular}[c]{@{}l@{}}0.4304\end{tabular}\\
Escape Velocity ($km/h$)
& Loglogistic 
&  \begin{tabular}[c]{@{}l@{}}$\mu$=2.10984\\$\sigma$=0.329425\end{tabular}
&  \begin{tabular}[c]{@{}l@{}} 0.7718\end{tabular}
& $t$ location-scale 
&  \begin{tabular}[c]{@{}l@{}}$\mu$= 8.24903\\$\sigma$= 2.61085\\$\nu$ = 1.15837\end{tabular}
&  \begin{tabular}[c]{@{}l@{}}0.7262\end{tabular} \\
\toprule
\textbf{Characteristics in Carme}&\textbf{Best-fit}&\textbf{Parameters}&\textbf{$p$-value}&\textbf{Best-fit}&\textbf{Parameters}&\textbf{$p$-value}\\
        \toprule
Equatorial Radius ($km$) & Loglogistic 
&  \begin{tabular}[c]{@{}l@{}} $\mu$=0.500116\\$\sigma$=0.345753\end{tabular}
&  \begin{tabular}[c]{@{}l@{}}0.6654\end{tabular}
& $t$ location-scale 
&  \begin{tabular}[c]{@{}l@{}}$\mu$ = 1.65708\\ $\sigma$ = 0.440683\\ $\nu$= 1.14501\end{tabular}
&  \begin{tabular}[c]{@{}l@{}}0.7119\end{tabular}\\

Surface Gravity ($10^{-2}m/s^{2}$)
& Loglogistic
&  \begin{tabular}[c]{@{}l@{}}$\mu$=-2.08871 \\$\sigma$=0.319434\end{tabular}
&  \begin{tabular}[c]{@{}l@{}}0.0584\end{tabular}
 &null & null & null \\

Mass ($10^{13}kg$) & Loglogistic
&  \begin{tabular}[c]{@{}l@{}} $\mu$=1.70257\\$\sigma$=0.956765\end{tabular}
&  \begin{tabular}[c]{@{}l@{}} 0.5776\end{tabular}
 & null & null  & null  \\

Escape Velocity($km/h$) 
& \begin{tabular}[c]{@{}l@{}}$t$ location-scale\\(Loglogistic)\end{tabular}   
&  \begin{tabular}[c]{@{}l@{}}$\mu$=7.08058\\$\sigma$=1.51796\\$\nu$=0.993412\\($\mu$=2.01252\\$\sigma$=0.30989) \end{tabular}
&  \begin{tabular}[c]{@{}l@{}}0.5996\\(0.5222)\end{tabular}
& $t$ location-scale 
&  \begin{tabular}[c]{@{}l@{}}$\mu$ = 7.32997\\ $\sigma$ = 1.58419\\ $\nu$= 1.06821\end{tabular}
&  \begin{tabular}[c]{@{}l@{}} 0.6619\end{tabular}\\
 \toprule
\textbf{Characteristics in Pasiphae}&\textbf{Best-fit}&\textbf{Parameters}&\textbf{$p$-value}&\textbf{Best-fit}&\textbf{Parameters}&\textbf{$p$-value}\\
        \toprule
Equatorial Radius ($km$) 
& Loglogistic 
&  \begin{tabular}[c]{@{}l@{}}$\mu$=0.802683\\$\sigma$= 0.59951\end{tabular}
&  \begin{tabular}[c]{@{}l@{}} 0.6575\end{tabular}
 & Weibull 
&  \begin{tabular}[c]{@{}l@{}} $A$= 3.6978\\ $B$ = 0.781648 \end{tabular}
&  \begin{tabular}[c]{@{}l@{}} 0.0859\end{tabular}\\

Surface Gravity ($10^{-2}m/s^{2}$) 
&Loglogistic
&  \begin{tabular}[c]{@{}l@{}}$\mu$=-1.68323\\$\sigma$= 0.555284\end{tabular}
&  \begin{tabular}[c]{@{}l@{}} 0.4431\end{tabular}
& Weibull 
&  \begin{tabular}[c]{@{}l@{}}$A$= 3.09876\\ $B$ = 0.843794 \end{tabular}
&  \begin{tabular}[c]{@{}l@{}} 0.0594\end{tabular} \\

Mass ($10^{13}kg$) 
& Inverse Gaussian 
&  \begin{tabular}[c]{@{}l@{}}$\mu$=3418.74\\$\lambda $=4.03647 \end{tabular}
&  \begin{tabular}[c]{@{}l@{}}0.8481 \end{tabular}
& Weibull 
&  \begin{tabular}[c]{@{}l@{}}$A$= 72.5121\\ $B$ = 0.266585 \end{tabular}
&  \begin{tabular}[c]{@{}l@{}} 0.1096\end{tabular}  \\

Escape Velocity ($km/h$)
& Loglogistic 
&  \begin{tabular}[c]{@{}l@{}}$\mu$=2.32956\\$\sigma$=0.570971 \end{tabular}
&  \begin{tabular}[c]{@{}l@{}}0.5832 \end{tabular}
& Weibull 
&  \begin{tabular}[c]{@{}l@{}}$A$= 17.0302\\ $B$ = 0.807236 \end{tabular}
&  \begin{tabular}[c]{@{}l@{}}0.0722 \end{tabular} \\
 \toprule
\end{tabular}
}
 \label{tab3-5} 
\end{table}

%\end{landscape}
As shown in Table 5, all four physical characteristics in the Ananke group follow a log-logistic distribution in this paper, and each of their $p$-values is larger than the corresponding $p$-value in [2]. In the Carme group, in addition to the escape velocity following a $t$ location-scale distribution or log-logistic distribution, the other three physical characteristics follow log-logistic distributions, while for the surface gravity and mass, the null hypothesis in [2] is rejected, and the remaining two physical characteristics follow a $t$ location-scale distribution. For the Pasiphae group, in addition to the mass following the inverse Gaussian distribution, the other three physical characteristics  follow the log-logistic distribution; the four physical characteristics all follow the Weibull distribution in [2]. The $p$-value of the distribution of the combined distribution is much larger than in [2]. For example, the $p$-value of the equatorial radius is 0.6575, while the $p$-value is only 0.0859 in [2]. 

We know that the log-logistic distribution is often used to analyze survival data, and its shape is similar to that of the log-normal and Weibull distributions [23]-[25]. The logarithmic log-normal distribution is also known as the normal distribution (or Gaussian distribution) [26], and even a log-logistic distribution can generate a Weibull distribution [27]. Therefore, it can be determined that although the irregular satellites have been regrouped and the number has increased, they follow similar shape distributions, as shown in both the present paper (Figures 1-3) and [2].

%Figure 1
\begin{figure}[htb]
\centering
\subfigure[]{
\begin{minipage}{18cm}
\centering
\subfigbottomskip=1pt 
\subfigcapskip=-5pt 
\includegraphics[width=9.5cm]{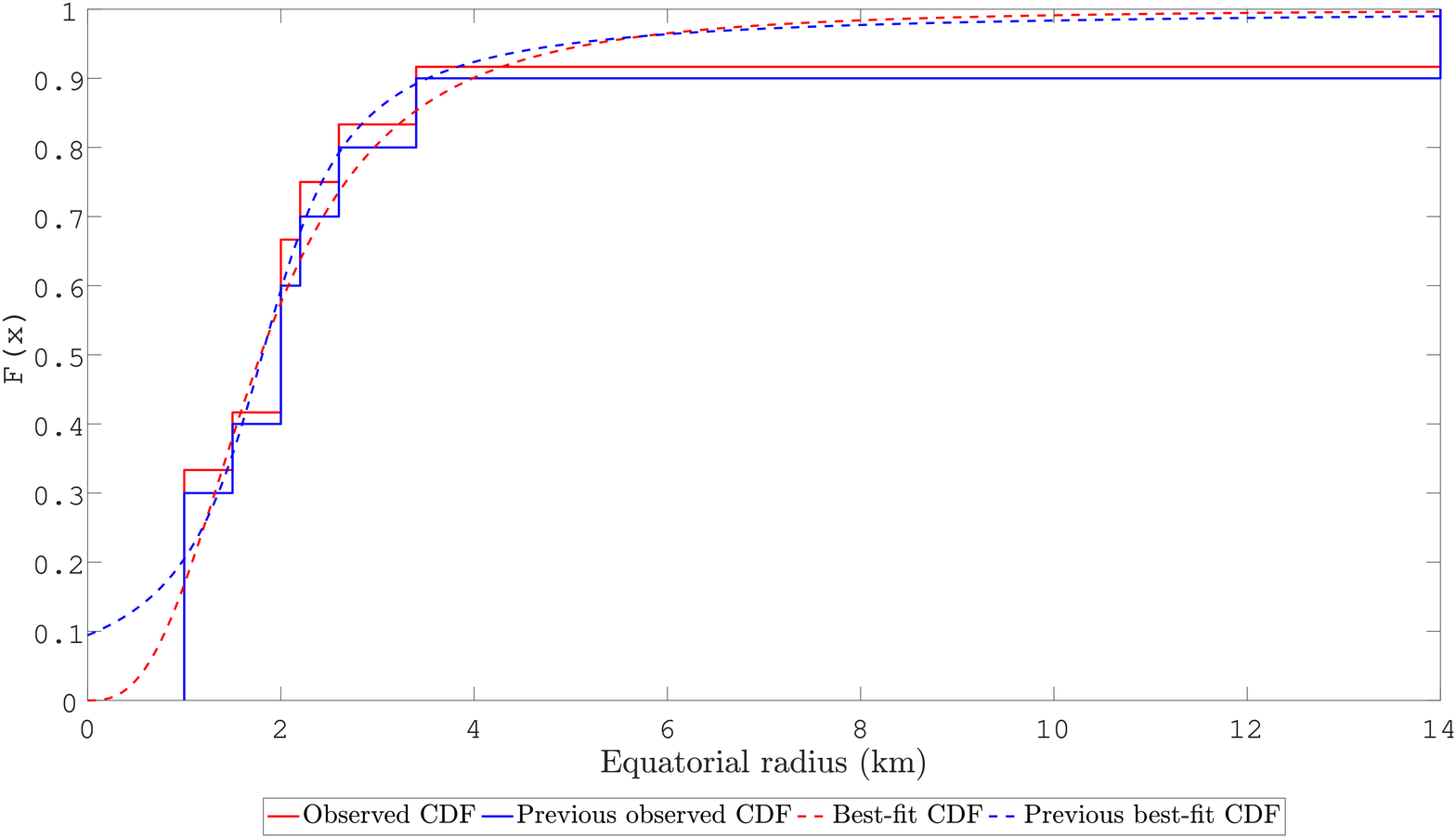}
\end{minipage}%
}\\
\subfigure[]{
\begin{minipage}{18cm}
\centering
\includegraphics[width=9.5cm]{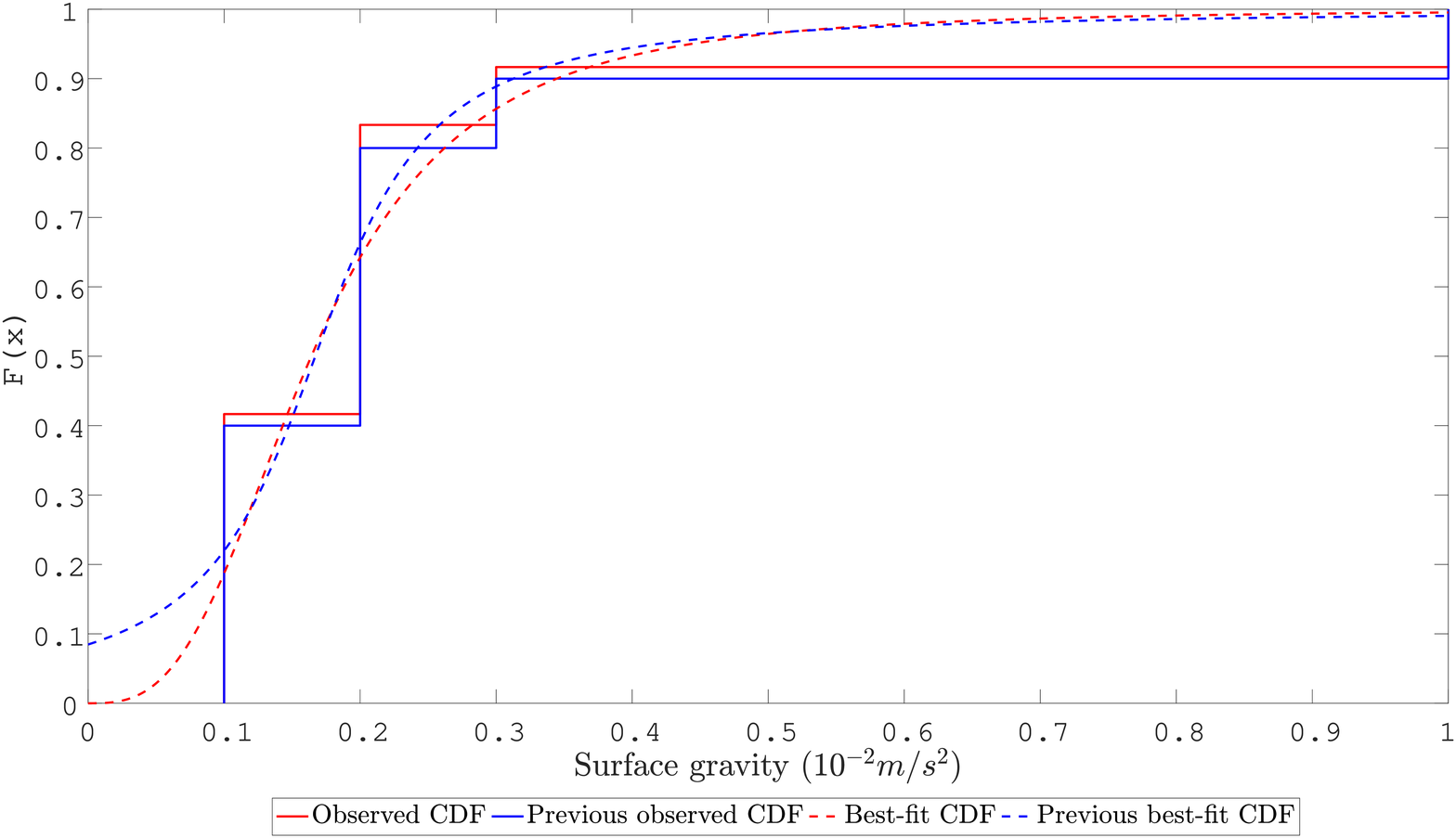}
\end{minipage}
}\\
\subfigure[]{
\begin{minipage}{18cm}
\centering
\includegraphics[width=9.5cm]{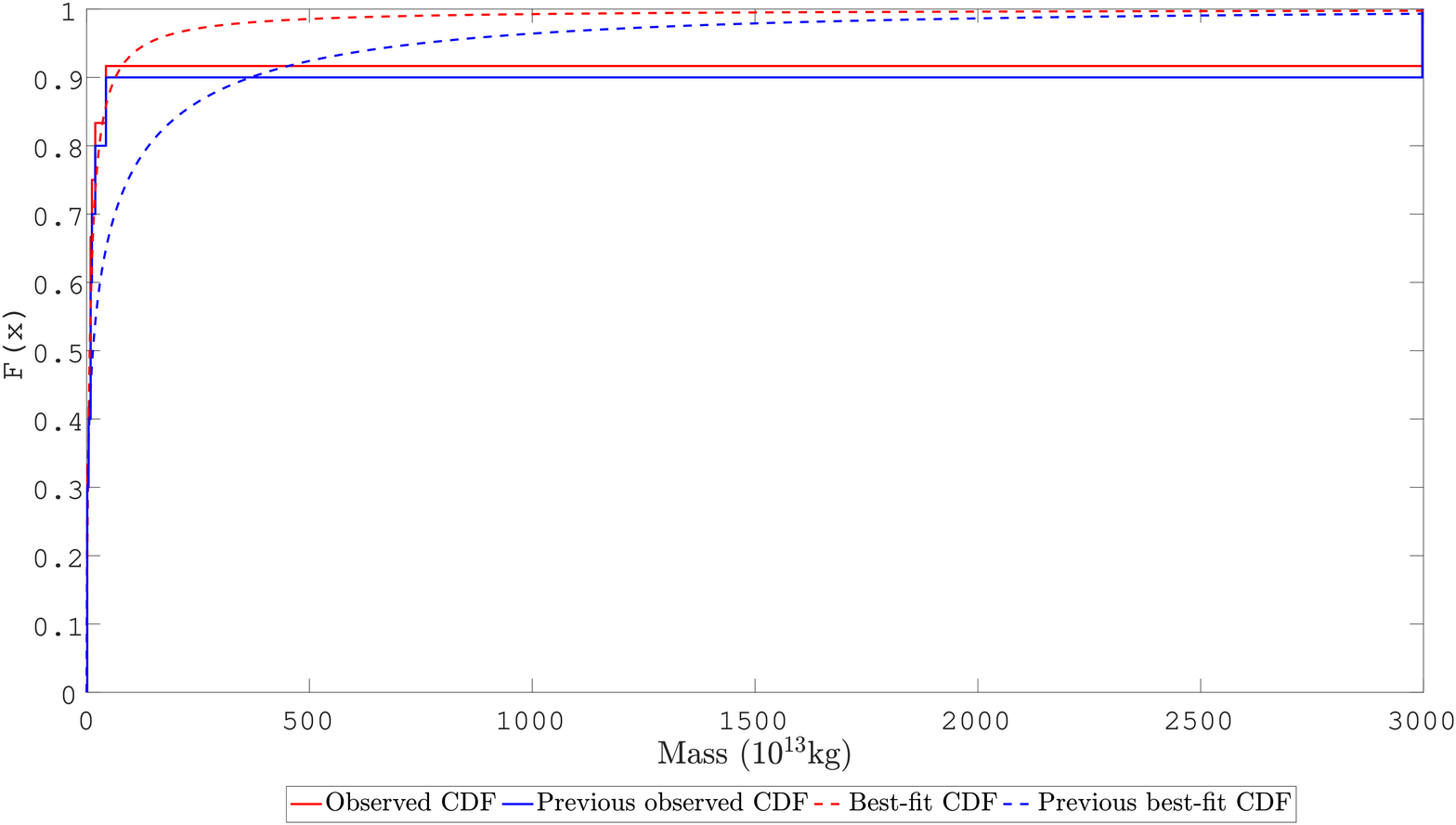}
\end{minipage}%
}\\
\subfigure[]{
\begin{minipage}{18cm}
\centering
\includegraphics[width=9.5cm]{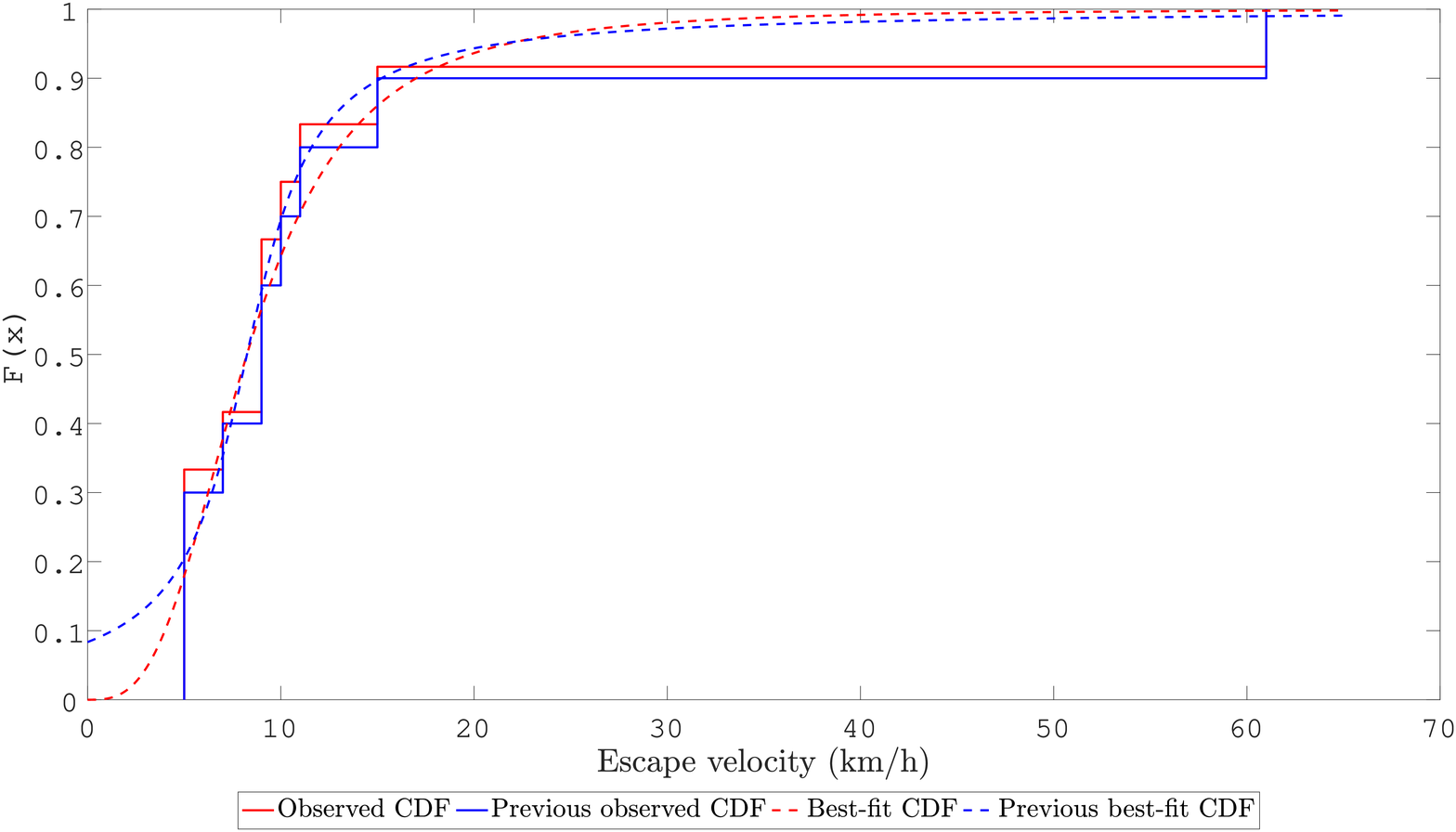}
\end{minipage}
}
\caption{(a), (b), (c) and (d) are the best-fit CDFs and the observed CDF of the current distributions and the previous distributions in [2] of the physical characteristics in the Ananke group, respectively.}
\label{fig21}
\end{figure}

\clearpage
%Figure 2
\begin{figure}[htb]
\subfigbottomskip=10pt 
\subfigcapskip=-62pt 
\subfigure[]{
\begin{minipage}{18cm}
\centering{\includegraphics[width=9.5cm]{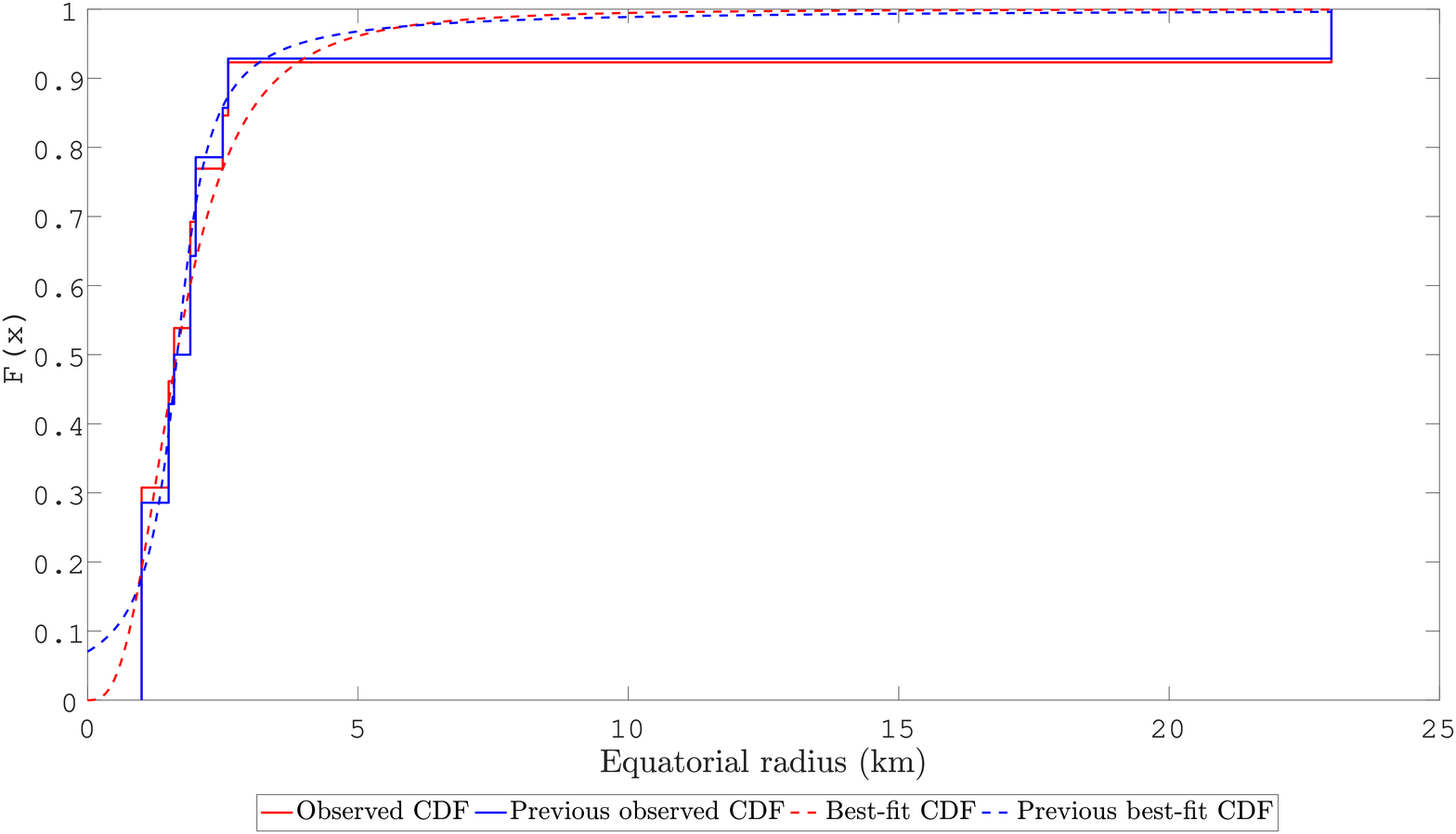}}
\vspace {22mm}
\end{minipage}%
}\\
\subfigure[]{
\begin{minipage}{18cm}
\centering
\includegraphics[width=9.5cm]{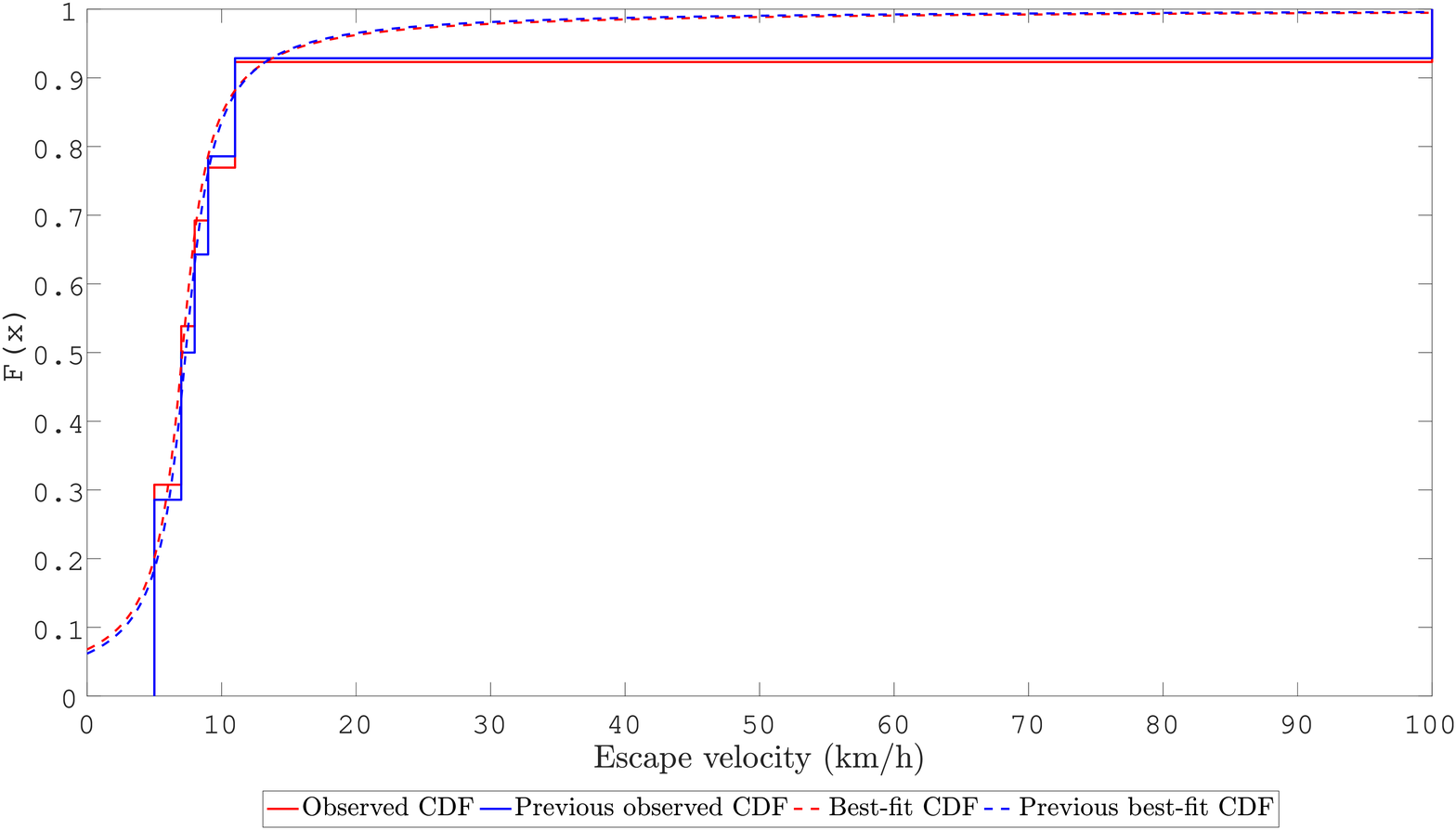}
\vspace {22mm}
\end{minipage}
}
\caption{(a) and (b) are the best-fit CDFs and the observed CDF of the current distributions and the previous distributions in [2] of the physical characteristics in the Carme group, respectively.}
\label{fig22}
\end{figure}

%Figure 3

\begin{figure}[htb]
\subfigure[]{
\begin{minipage}{18cm}
\centering
\subfigbottomskip=1pt 
\subfigcapskip=-5pt 
\includegraphics[width=9.5cm]{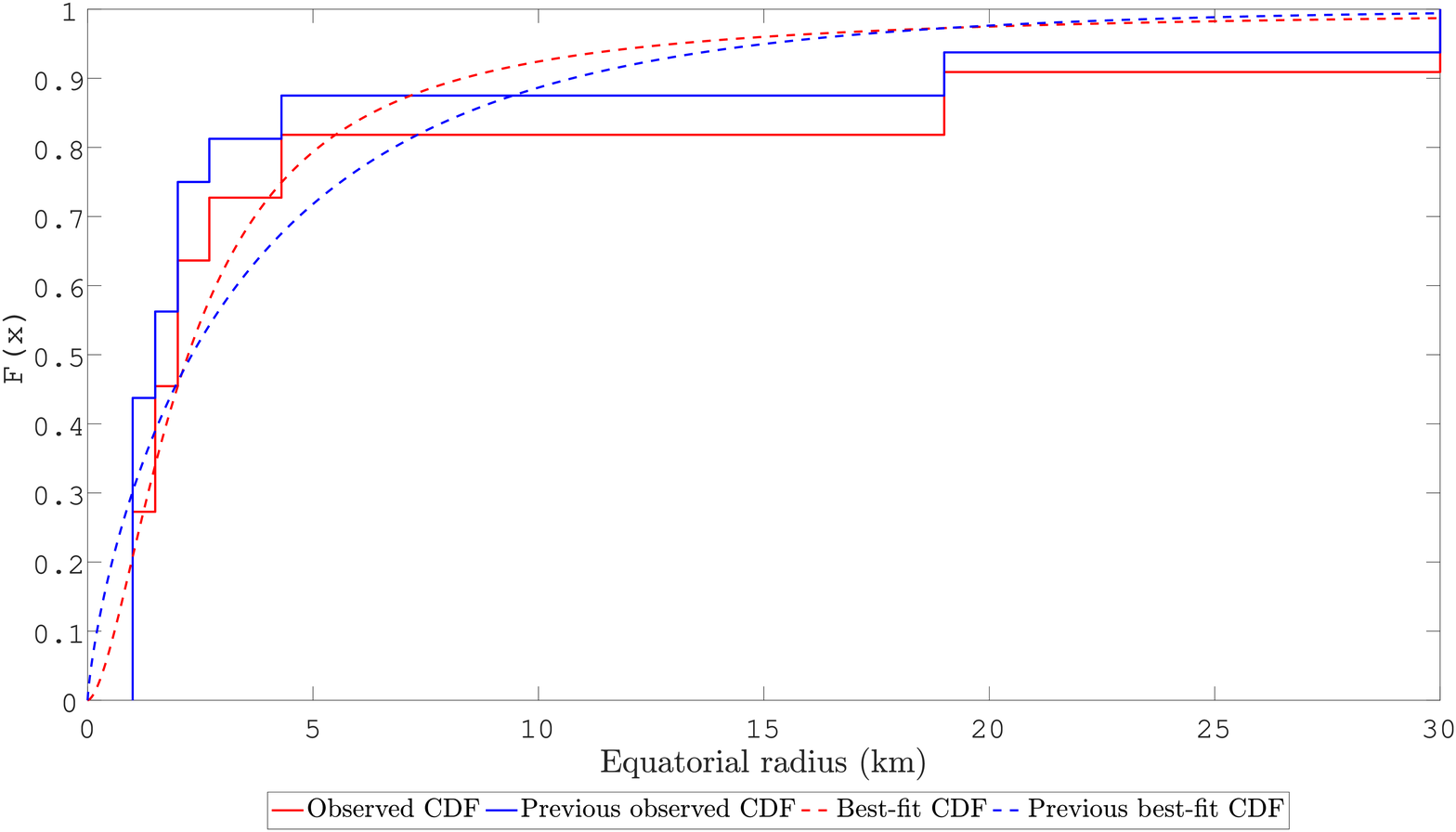}
\vspace {0mm}%
\end{minipage}%
}\\
\subfigure[]{
\begin{minipage}{18cm}
\centering
\includegraphics[width=9.5cm]{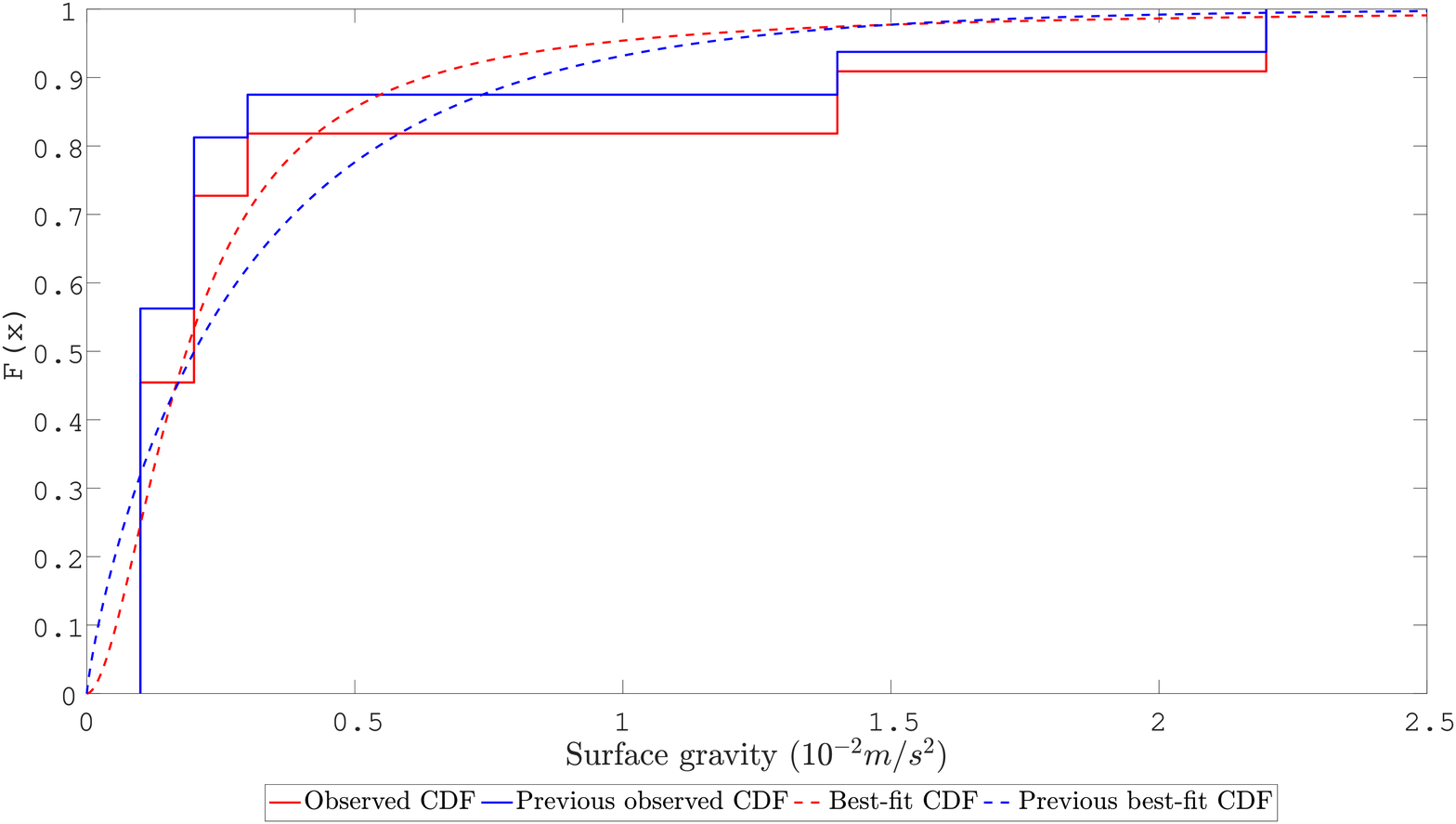}
\vspace {0mm}%
\end{minipage}
}\\
\subfigure[]{
\begin{minipage}{18cm}
\centering
\includegraphics[width=9.5cm]{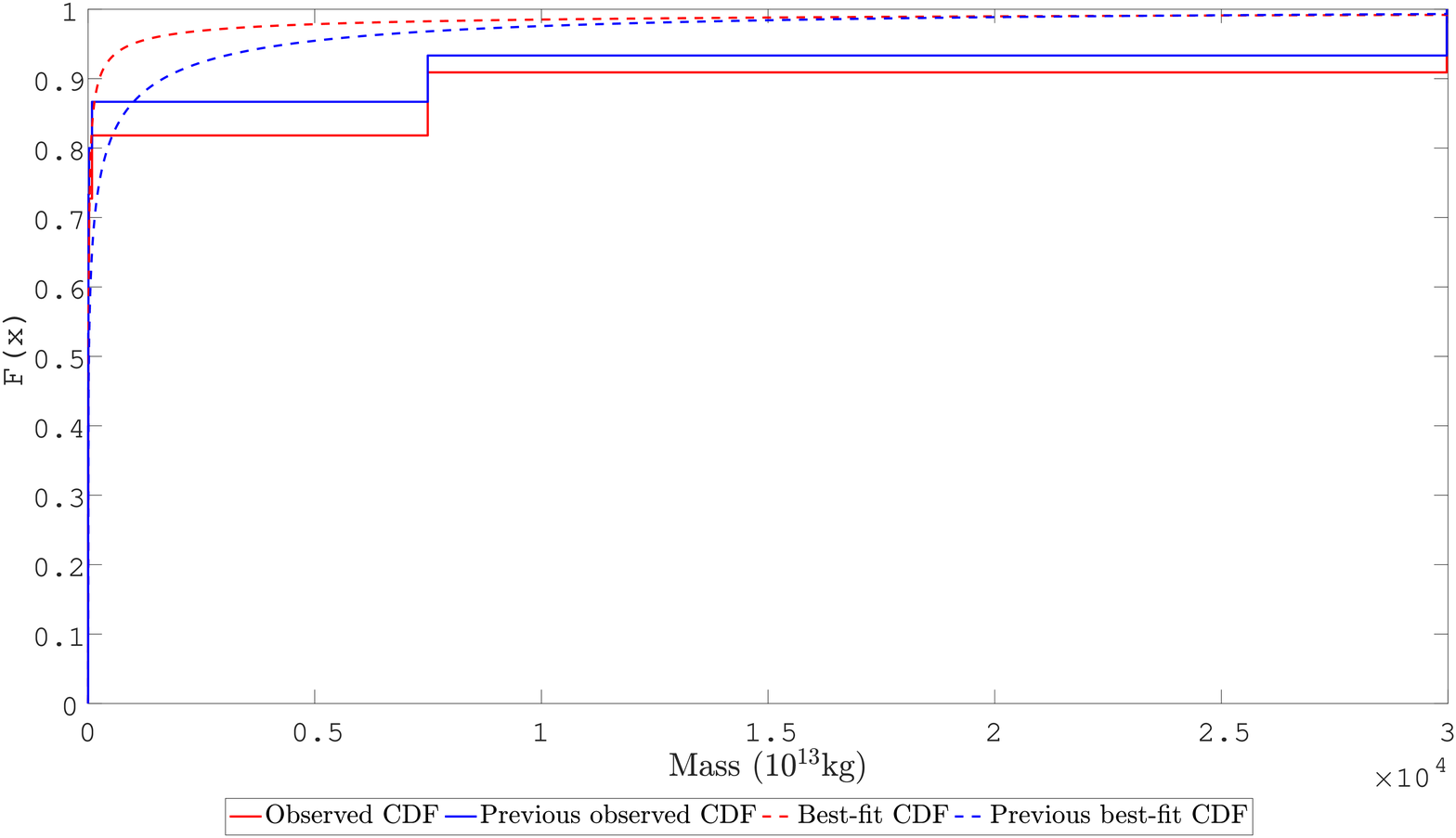}
\end{minipage}%
}\\
\subfigure[]{
\begin{minipage}{18cm}
\centering
\includegraphics[width=9.5cm]{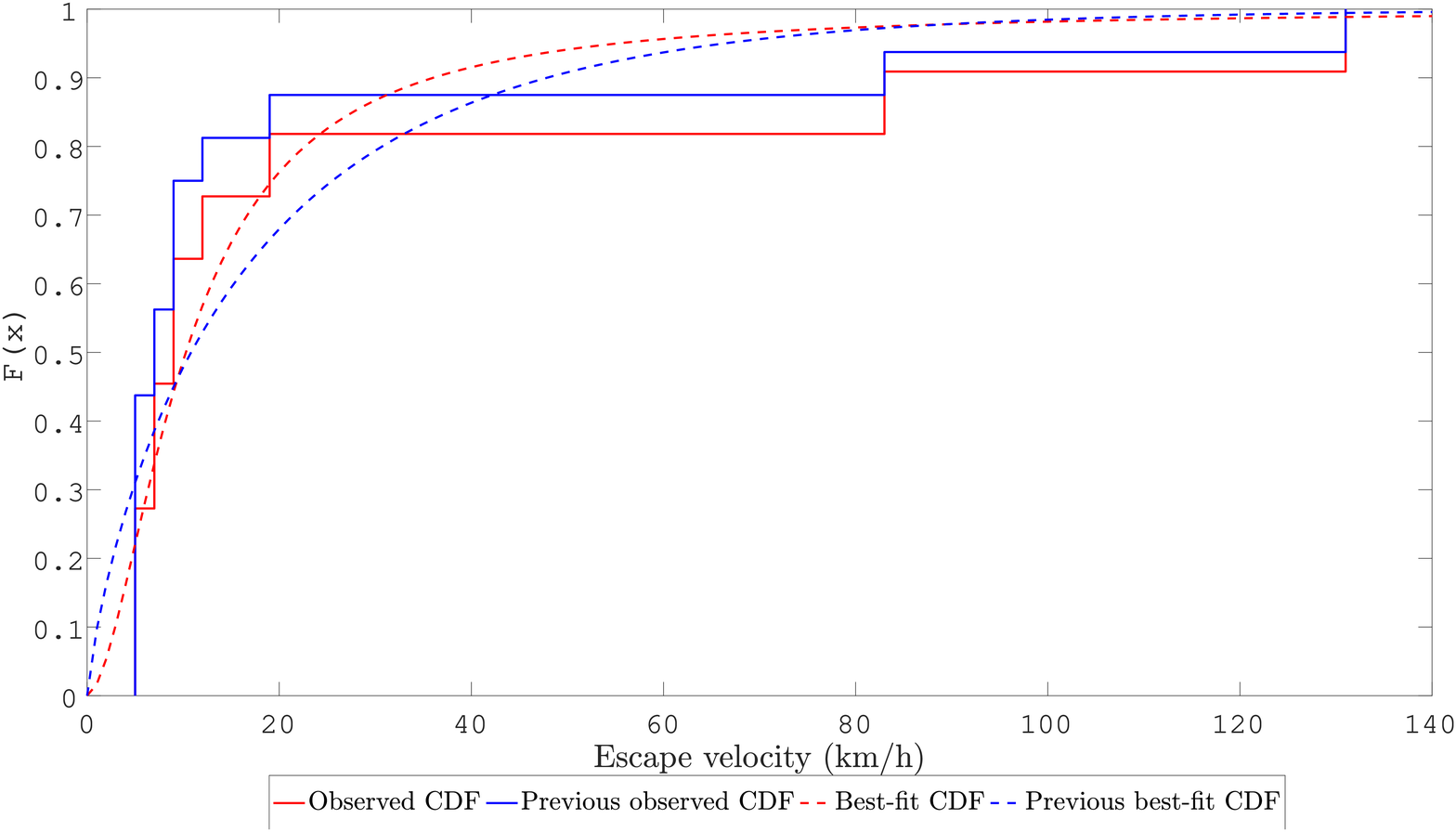}
\end{minipage}
}
\caption{(a), (b), (c) and (d) are the best-fit CDFs and the observed CDF of the current distributions and previous distributions in [2] of the physical characteristics in the Pasiphae group, respectively.}
\label{fig23}
\end{figure}

\section{Verification of the rationality of the statistical inference results}

\subsection{Ananke group}

According to the observation data of $V$ and $R$, if their statistically predicted distributions are $ d_ {pre, V} $ and $ d_ {pre, R} $, respectively, note that there is a nonlinear relationship $ V = 4 \pi R ^ 3/3 $ between these two characteristics. The analytical distribution $d_{ana, V}$ of $V$ can then be analytically derived from the statistically predicted distribution $d_{pre, R}$ of $R$. The same method can also be applied to obtain the analytical distribution $ d_{ana, R} $ of $R$ according to the statistically predicted distribution $ d_{pre, V} $ of $ V $. Therefore, if $ d_{pre, V} $ and $ d_{ana, V} $ (or $ d_{pre, R} $ and $ d_{ana, R} $) have the same mathematical expression or have matching PDF curves, then the statistical prediction results in Section 3 have very high reliability from the perspective of strict analytical derivation.

Let the statistically predicted PDF of $R$ be $f_{pre,R}$. Note that the derivative of $R$ is
\begin{equation}
\begin{aligned}
R^{'}=\frac{[3/(4\pi)]^{\frac{1}{3}}}{3}V^{-\frac{2}{3}}.
\label{eq24}
\end{aligned}
\end{equation}
The PDF of $V$ can then be rewritten as 
\begin{equation}
\begin{aligned}
f_{ana,V}\left ( V;\mu ,\sigma \right )
=f_{pre,R}\left (  \left (\frac{3V}{4\pi } \right )^{\frac{1}{3}};\mu,\sigma \right )\frac{\left [ 3/(4\pi)  \right ]^{\frac{1}{3}}}{3}V^{-\frac{2}{3}}.\label{eq25}
\end{aligned}
\end{equation}

Table 1 shows that $V$ and $R$ follow log-logistic distributions with parameters (3.17586, 1.10596) and (0.582917, 0.364002), respectively. The PDF of $V$ can be obtained analytically by the statistically predicted PDF of $R$ as follows:
\begin{equation}
\begin{aligned}
f_{ana,V}(V;\mu,\sigma )=\frac{0.916e^{2.747\ln\left(0.620V^{\frac{1}{3}}\right)-1.601}}{V\left [1+e^{2.747\ln\left(0.620V^{\frac{1}{3}}\right)-1.601} \right ]^{2}}.
\label{eq27}
\end{aligned}
\end{equation}
From equation (2), the statistically predicted PDF of $V$ can be written as
\begin{equation}
\begin{aligned}
f_{pre,V}(V;\mu ,\sigma )=\frac{1}{\sigma V }\frac{e^{\frac{\ln V-\mu }{\sigma }}}{\left[1+e^{\frac{\ln V-\mu }{\sigma }}\right]^2}=\frac{0.904e^{0.904\ln V-2.872}}{V\left(1+e^{0.904\ln V-2.872}\right)^2}.
\label{eq26}
\end{aligned}
\end{equation}

As shown in Figure 4, the analytically derived $f_{ana, V}$ is in good agreement with the statistically predicted $f_{pre, V}$.

%radius to volume 
 %figure 4
   \begin{figure}[htb]
  \center
\begin{minipage}[htb]{165mm}
\centerline{\includegraphics[scale=0.6]{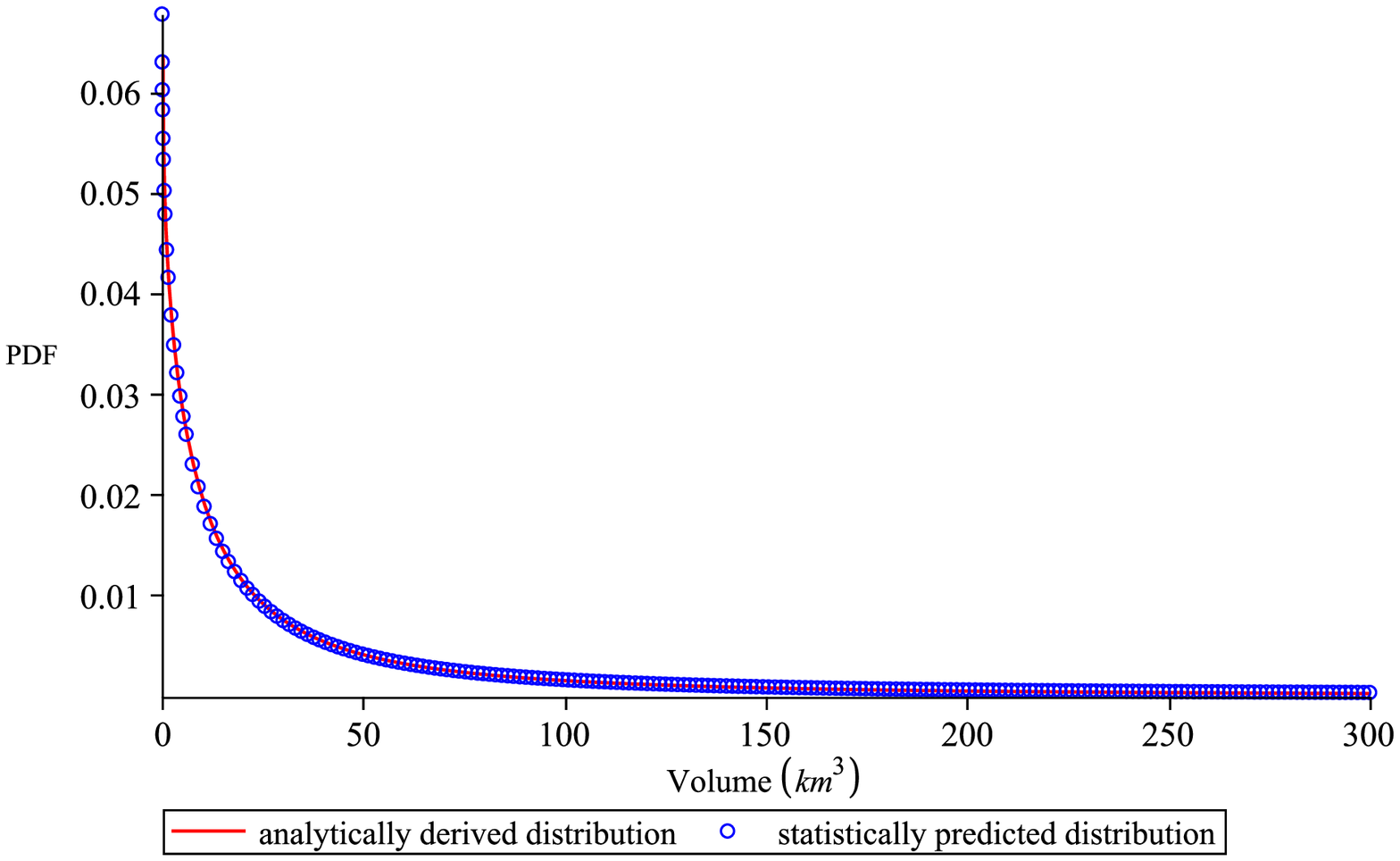}}
\vspace {-1mm}
\caption{Comparison of the PDF curves between the statistically predicted distribution and the analytically derived distribution based on $R$ in the Ananke group}
\label{fig24}
\end{minipage}
\end{figure}

%volume to radius 
In the same way, based on $R=\left ( 3V/4\pi \right )^{\frac{1}{3}}$, we can also analytically derive the distribution of $R$ from the statistically predicted distribution of $V$ as follows:
\begin{equation}
\begin{aligned}
f_{ana,R}(R;\mu,\sigma)=4\pi R ^{2}f_{pre,V}\left ( \dfrac{4}{3}\pi R^{3};\mu,\sigma \right ).
\label{eq28}
\end{aligned}
\end{equation}
Thus, the PDF of $R$ is analytically obtained as follows:  
\begin{equation}
\begin{aligned}
f_{ana,R}(R;\mu,\sigma)=\frac{2.713e^{0.904\ln\left(4\pi R^3/3 \right )-2.872}}{R\left [ 1+e^{0.904\ln\left(4\pi R^3/3 \right )-2.872} \right ]^{2}}.
\label{eq210}
\end{aligned}
\end{equation}
Correspondingly, the statistically predicted PDF of $R$ can be written as
\begin{equation}
\begin{aligned}
f_{pre,R}\left ( R;\mu,\sigma  \right )=\frac{1}{\sigma R }\frac{e^{\frac{\ln R-\mu }{\sigma }}}{(1+e^{\frac{\ln R-\mu }{\sigma }})^2}=\frac{2.747e^{2.747\ln R-1.601}}{R\left( 1+e^{2.747\ln R-1.601} \right)^{2}}.
\label{eq29}
\end{aligned}
\end{equation}

The statistically predicted and analytically derived PDFs of $R$ are shown in Figure 5. These PDFs match well with each other.

 %figure 5
\begin{figure}[htb]
\begin{minipage}[htb]{165mm}
\vspace {2mm}
\centerline{\includegraphics[scale=0.6]{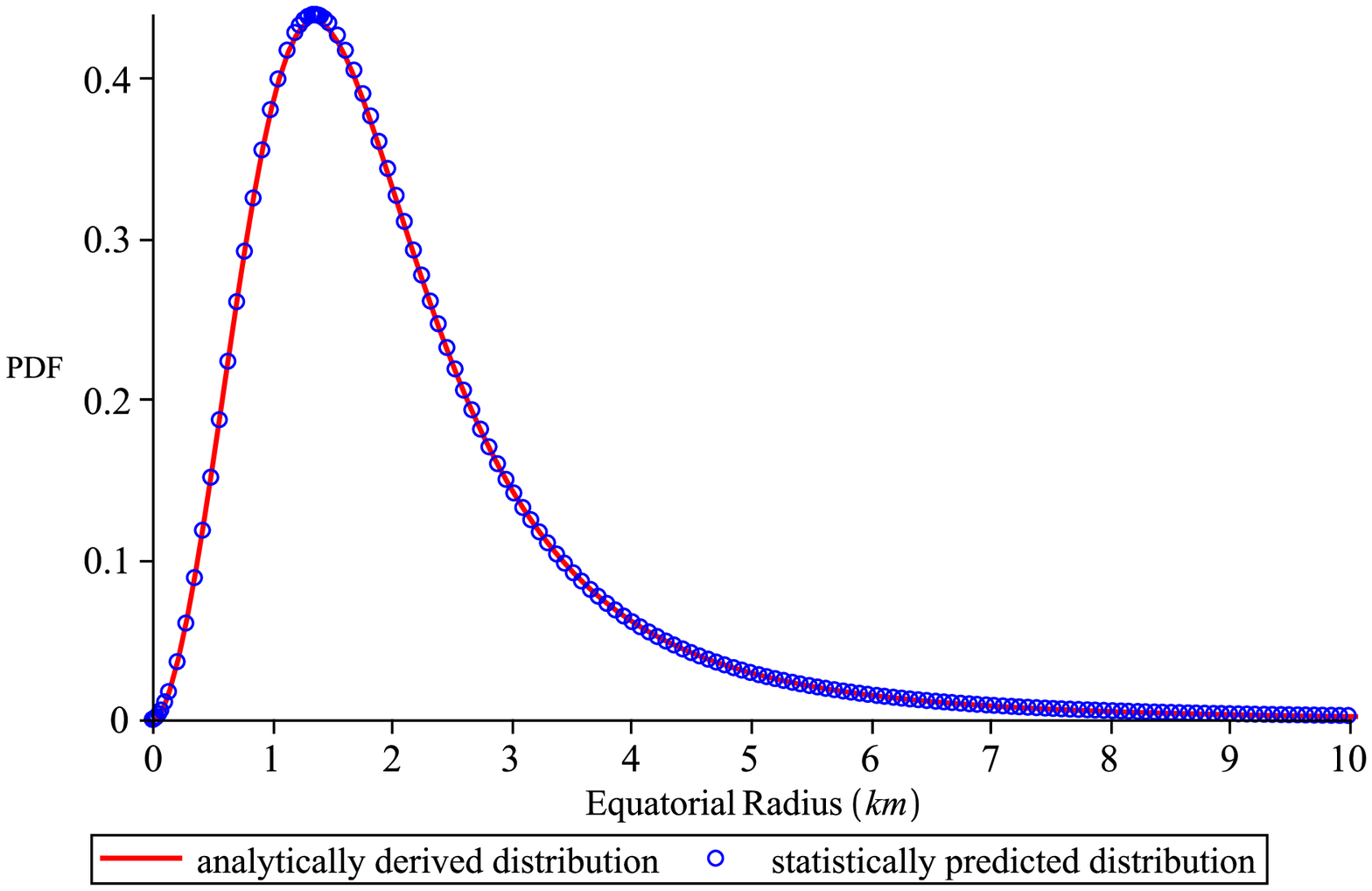}}
\vspace {-1mm}
\caption{Comparison of the PDF curves between the statistically predicted distribution and the analytically derived distribution based on $V$ in the Ananke group}
\label{fig25}
\end{minipage}
\end{figure}

%c2r
In addition, due to the linear relationship between $C$ and $R$, the PDF of $R$ derived from the distribution of $C$ by the aforementioned method must be consistent with the result of the statistically predicted PDF of $R$ (see Figure 6).  

 %figure 6
\begin{figure}[htb]
\center
\begin{minipage}[htb]{165mm}
\vspace {2mm}
\centerline{\includegraphics[scale=0.6]{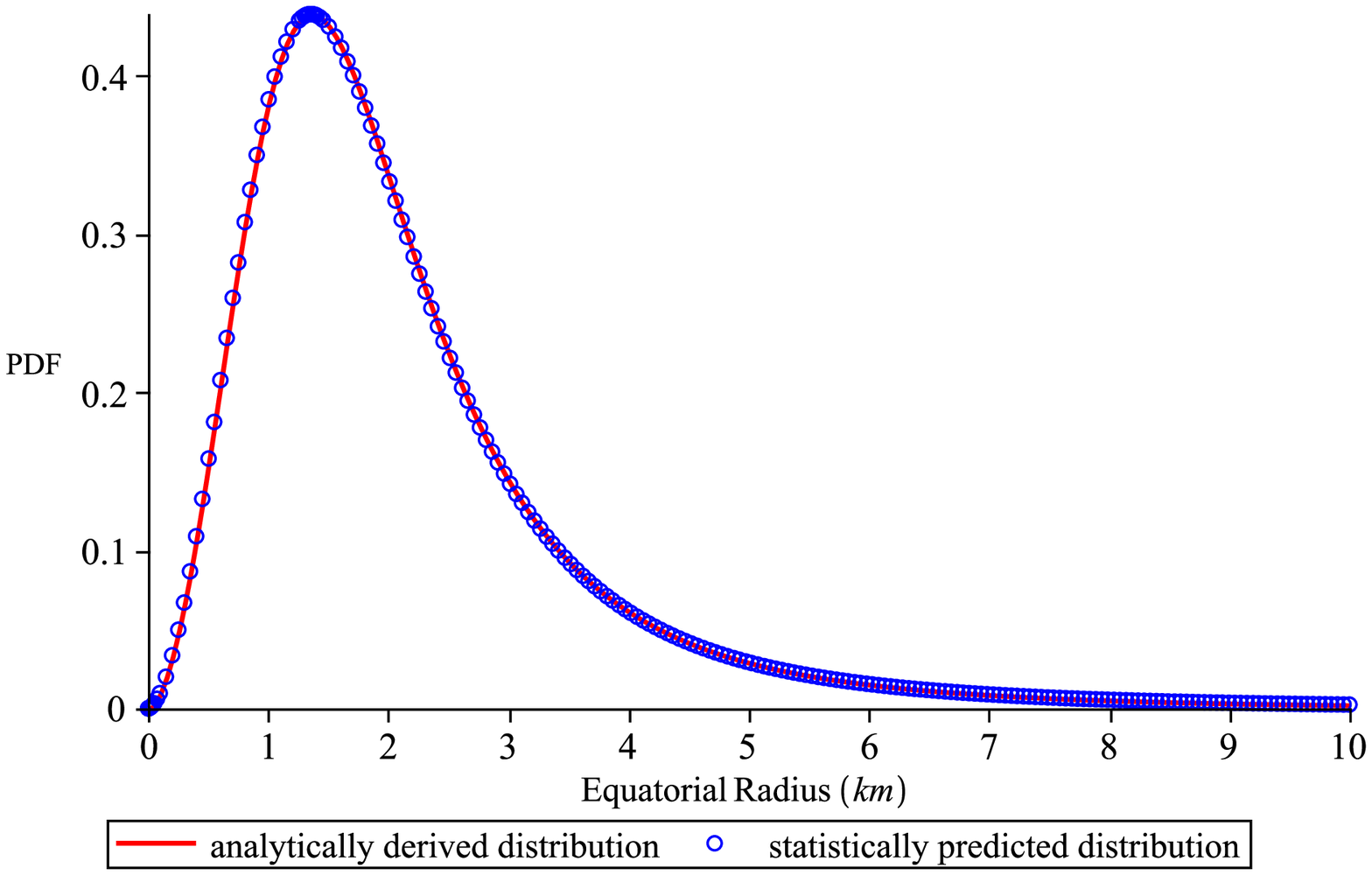}}
\vspace {-1mm}
\caption{Comparison of the PDF curves between the statistically predicted distribution and the analytically derived distribution based on $C$ in the Ananke group}
\label{fig26}
\end{minipage}
\end{figure}

Note that there is a mathematical relationship $R=\sqrt{S/4\pi }$ between $R$ and the surface area.
Let the statistically predicted PDF of $S$ be $f_{pre,S}$ and the analytically derived distribution of $R$ be $f_{ana,R}$. We then have
\begin{equation}
\begin{aligned}
f_{ana,R}\left ( R;\mu,\sigma \right )=8\pi R f_{pre,S}\left ( 4\pi R^{2};\mu,\sigma \right ).
\label{eq211}
\end{aligned}
\end{equation}
As shown in Table 1, both $R$ and $S$ follow log-logistic distributions with parameters (0.582917, 0.364002) and (3.69694, 0.727939), respectively.
The analytically derived PDF of $R$ then becomes
\begin{equation}
\begin{aligned}
f_{ana,R}(R;\mu,\sigma )=\frac{2.747e^{1.374\ln(12.566R^{2})-5.079}}{R\left [ 1+e^{1.374\ln(12.566R^{2})-5.079}\right ]^{2}}.
\label{eq212}
\end{aligned}
\end{equation}

Combining equations (9) and (11), Figure 7 shows the statistically predicted and analytically derived PDFs of $R$. They are in good agreement with each other.

%figure 7
   \begin{figure}[htb]
  \center
\begin{minipage}[htb]{165mm}
\vspace {2mm}
\centerline{\includegraphics[scale=0.6]{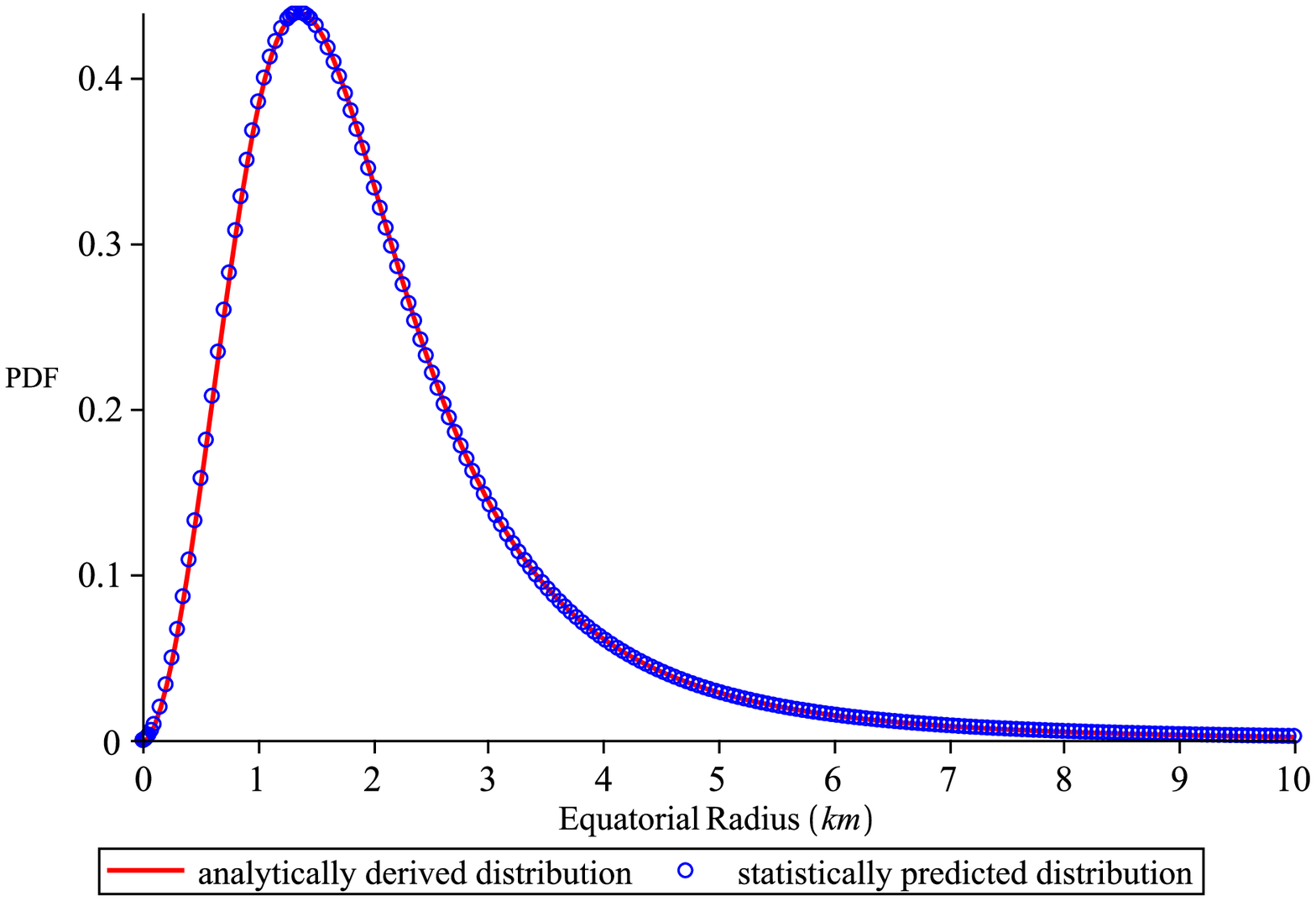}}
\vspace {-1mm}
\caption{Comparison of the PDF curves between the statistically predicted distribution and the analytically derived distribution based on $S$ in the Ananke group}
\label{fig27}
\end{minipage}
\end{figure}

Figures 5-7 show the PDF curves of $R$ obtained from the distributions of $V$, $C$ and $S$, respectively, and their shapes look extremely similar or even the same. The rationality of the results is verified to some extent from the perspective of the connection between different physical characteristics.

Similarly, the PDFs of $S$ can be analytically derived from the distribution of $R$. We then have
\begin{equation}
\begin{aligned}
f_{ana,S}(S;\mu ,\sigma )=\frac{1}{4\sqrt{\pi S}}f_{pre,R}\left ( \sqrt{\frac{S}{4\pi }} ;\mu ,\sigma \right )=\frac{1.374e^{2.747\ln\left(0.282\sqrt{S}\right)-1.601}}{S\left [1+ e^{2.747\ln(0.282\sqrt{S})-1.601}\right ]^{2}}.
\label{eq215}
\end{aligned}
\end{equation}
Note that the statistically predicted PDF of $S$ can be written as
\begin{equation}
\begin{aligned}
f_{pre,S}(S;\mu ,\sigma )=\frac{1}{\sigma S }\frac{e^{\frac{\ln S-\mu }{\sigma }}}{\left(1+e^{\frac{\ln S-\mu }{\sigma }} \right)^{2}}=\frac{1.374e^{1.374\ln S-5.079}}{S\left ( 1+e^{1.374\ln S-5.079} \right )^{2}}.
\label{eq214}
\end{aligned}
\end{equation}

These two PDFs are illustrated in Figure 8 and obviously match very well.
\clearpage
%figure 8
   \begin{figure}[htb]
 \center
\begin{minipage}[htb]{165mm}
\vspace {2mm}
\centerline{\includegraphics[scale=0.6]{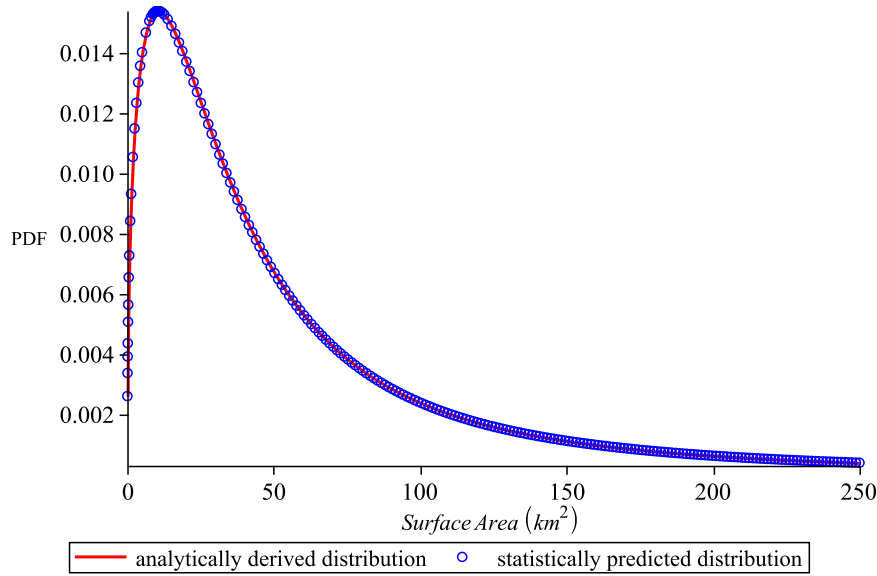}}
\vspace {-1mm}
\caption{Comparison of the PDF curves between the statistically predicted and analytically derived distributions based on $R$ in the Ananke group}
\label{fig28}
\end{minipage}
\end{figure}

\subsection{Carme group}
By using the same method, the comparison of the PDF curves from the perspective of statistical prediction and analytical derivation is shown in Figures 9 and 10. These two types of curves also agree well with each other.

%Figure 9
\begin{figure}[htb]
  \centering
  \subfigure[]{\label{fig:subfig:a1}
    \includegraphics[width=11cm]{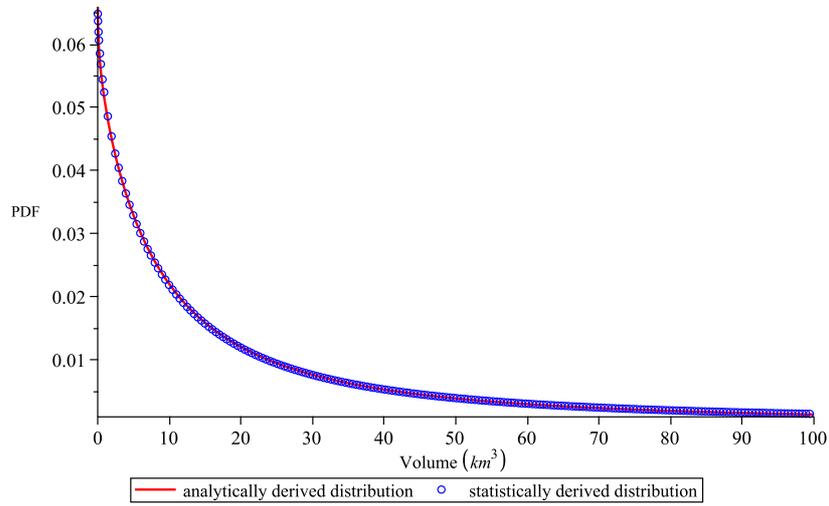}}
  \subfigure[]{\label{fig:subfig:b1}
    \includegraphics[width=11cm]{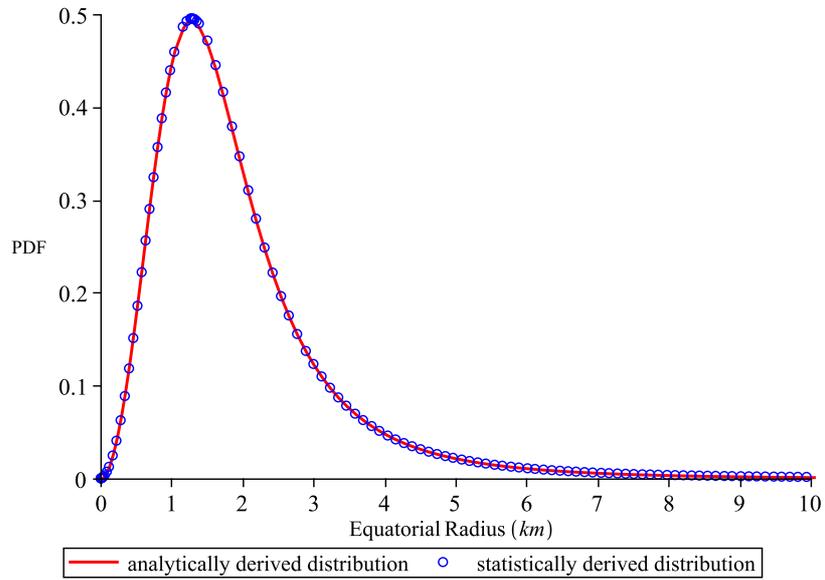}}
  \hspace{1in}
  \subfigure[]{\label{fig:subfig:c1}
    \includegraphics[width=11cm]{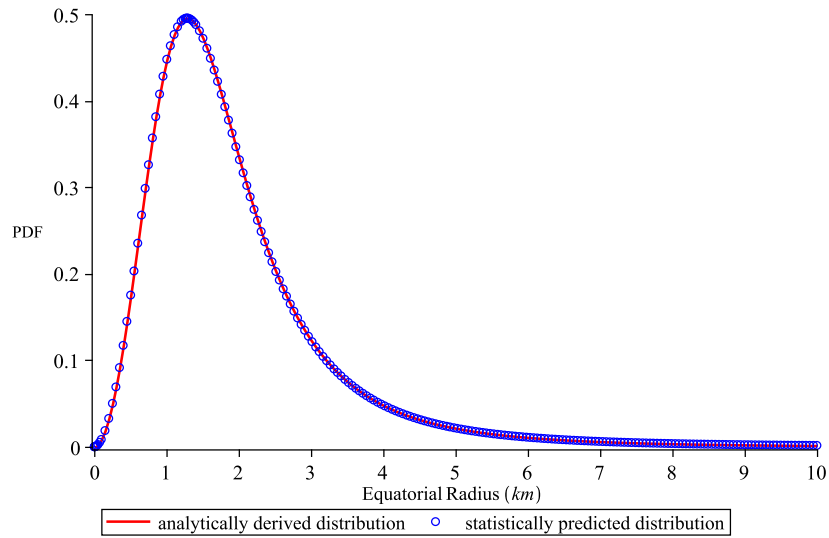}}

  \caption{(a) shows the statistically predicted and analytically derived PDF curves of $V$ based on $R$. (b) shows the statistically predicted and analytically derived PDF curves of $R$ based on $V$. (c) shows the statistically predicted and analytically derived PDF curves of $R$ based on $S$.}
  \label{fig:subfig}
  \label{fig29}
\end{figure}

%Figure 10
   \begin{figure}[htb]
  \center
\begin{minipage}[htb]{165mm}
\vspace {2mm}
\centerline{\includegraphics[scale=0.6]{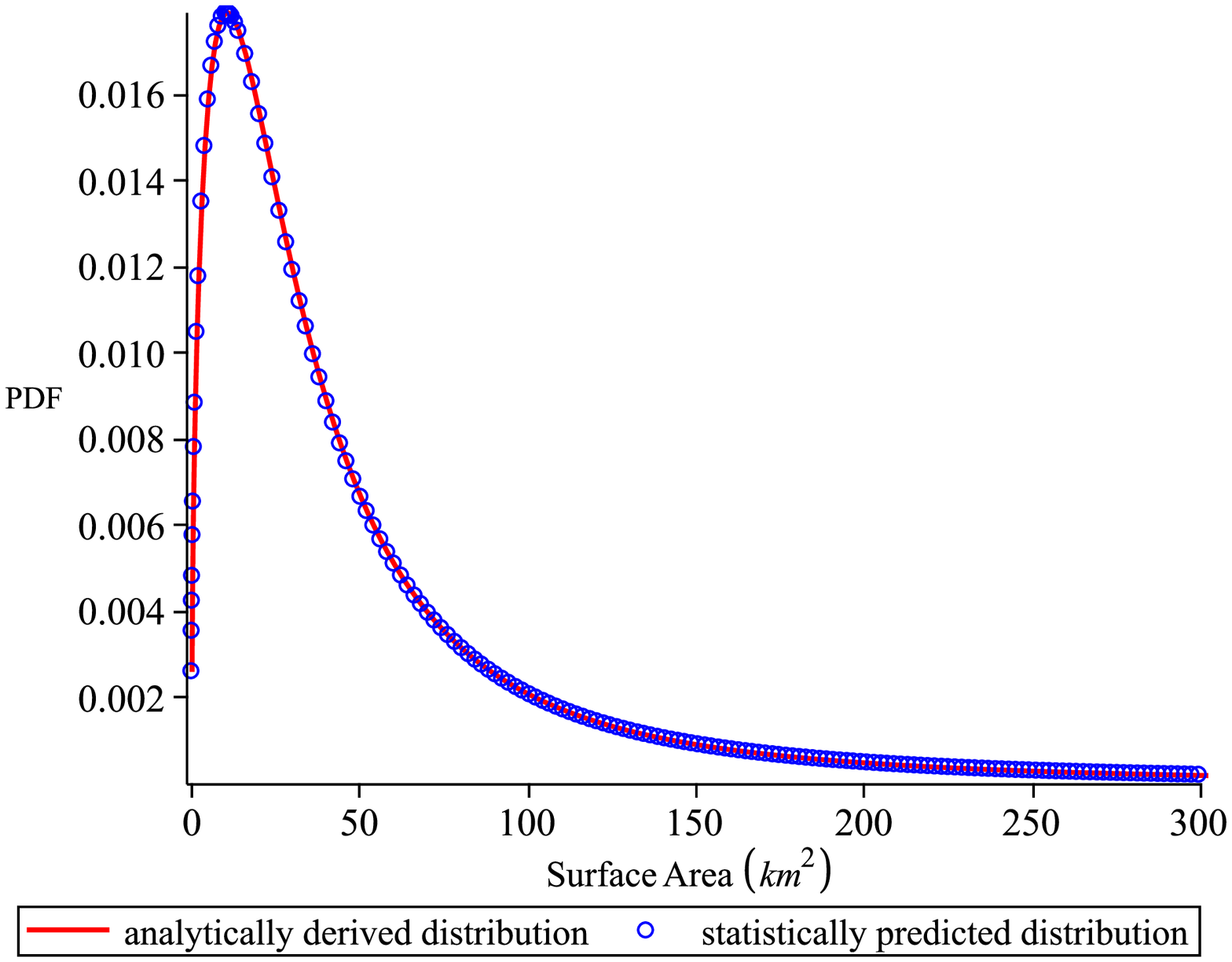}}
\vspace {-1mm}
\caption{Comparison of the PDF curves between the statistically predicted and analytically derived distributions in the Carme group}
\label{fig210}
\end{minipage}
\end{figure}

\subsection{Pasiphae group}

In this group, the previous methods are also used to compare the PDF curves from the perspective of statistical prediction and analytical derivation, as shown in Figures 11 and 12. However, although the two types of curves in this group have a certain degree of agreement, the effect of coincidence is not as good as in the Ananke group and the Carme group.

 %figure 11
   \begin{figure}[htb]
  \center
\begin{minipage}[htb]{165mm}
\vspace {2mm}
\centerline{\includegraphics[scale=0.7]{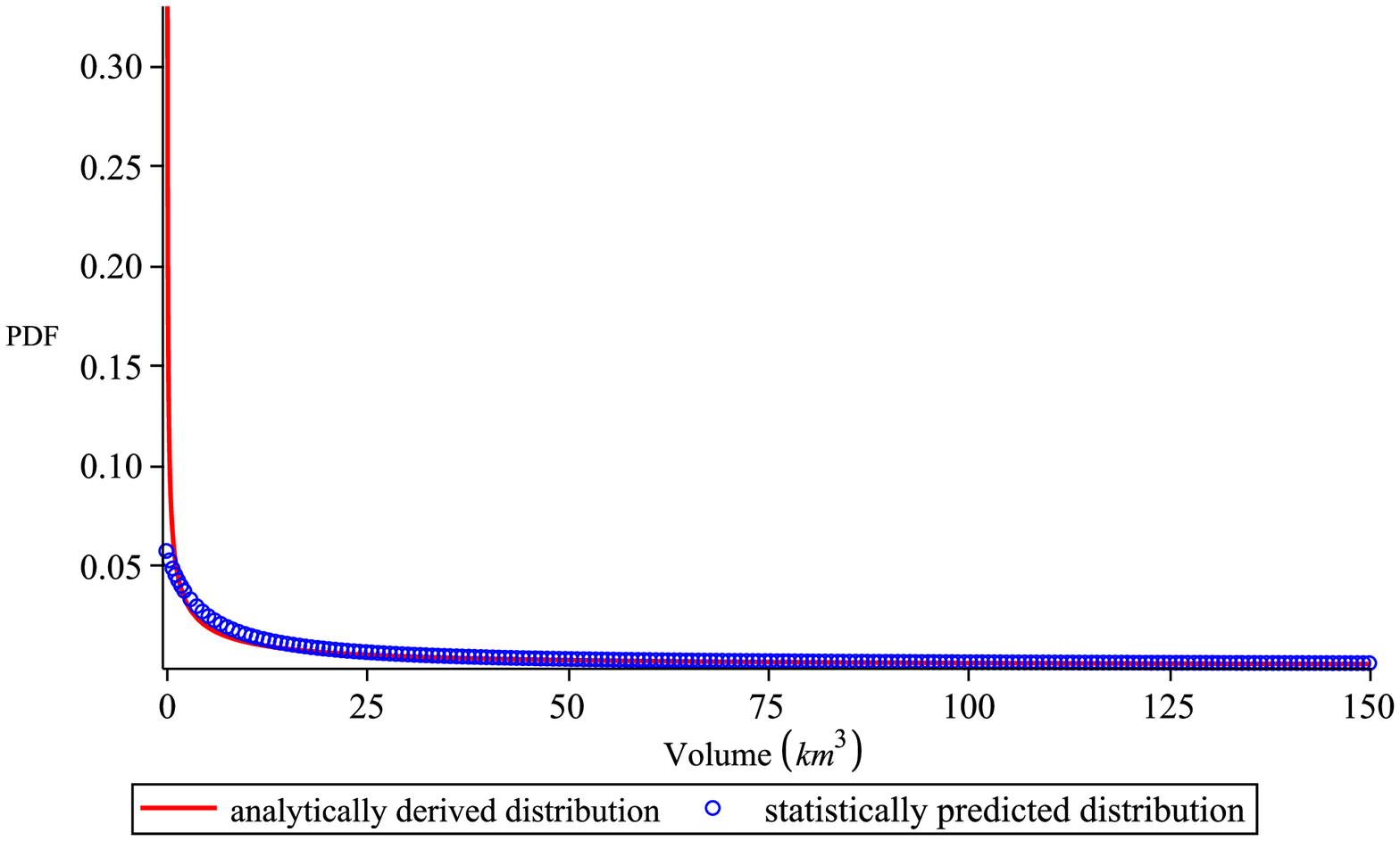}}
\vspace {-1mm}
\caption{Comparison of the PDF curves between the statistically predicted and analytically derived distributions in the Pasiphae group}
\label{fig211}
\end{minipage}
\end{figure}

 %figure 12
\begin{figure}[htb]
  \centering
  \subfigure[]{\label{fig:subfig:a1}
    \includegraphics[width=10cm]{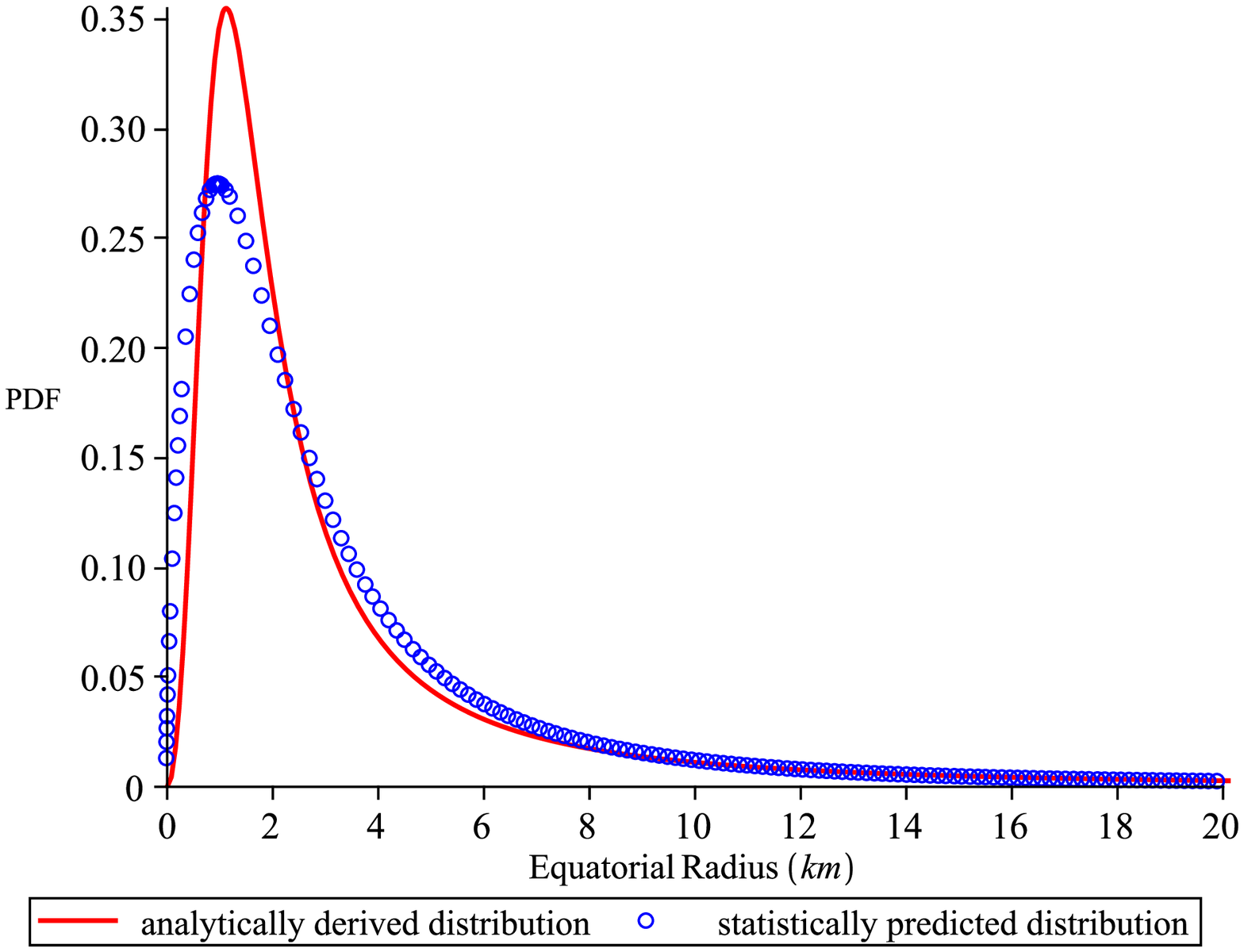}}
  \subfigure[]{\label{fig:subfig:b1}
    \includegraphics[width=10cm]{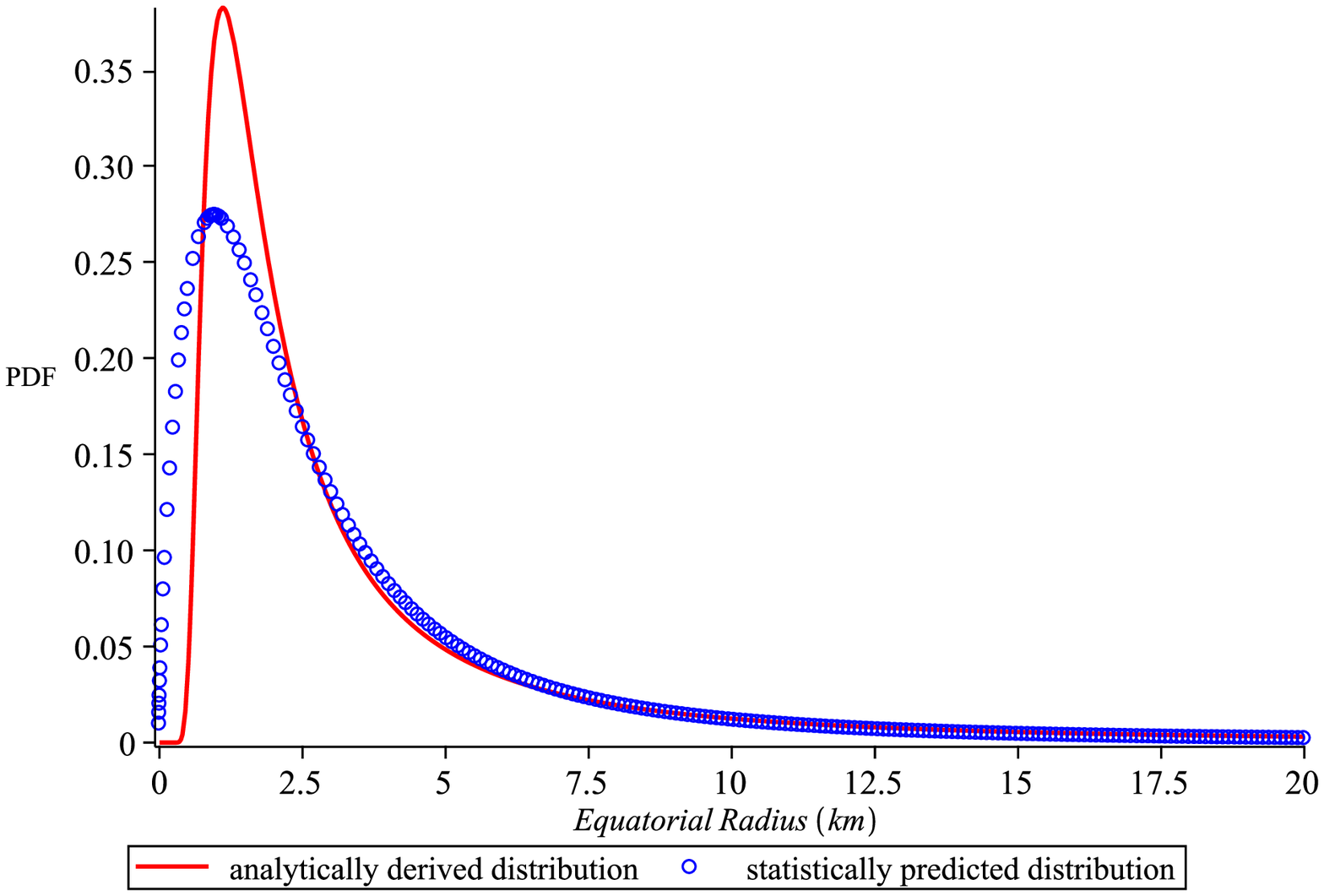}}
  \hspace{1in}
  \subfigure[]{\label{fig:subfig:c1}
    \includegraphics[width=10cm]{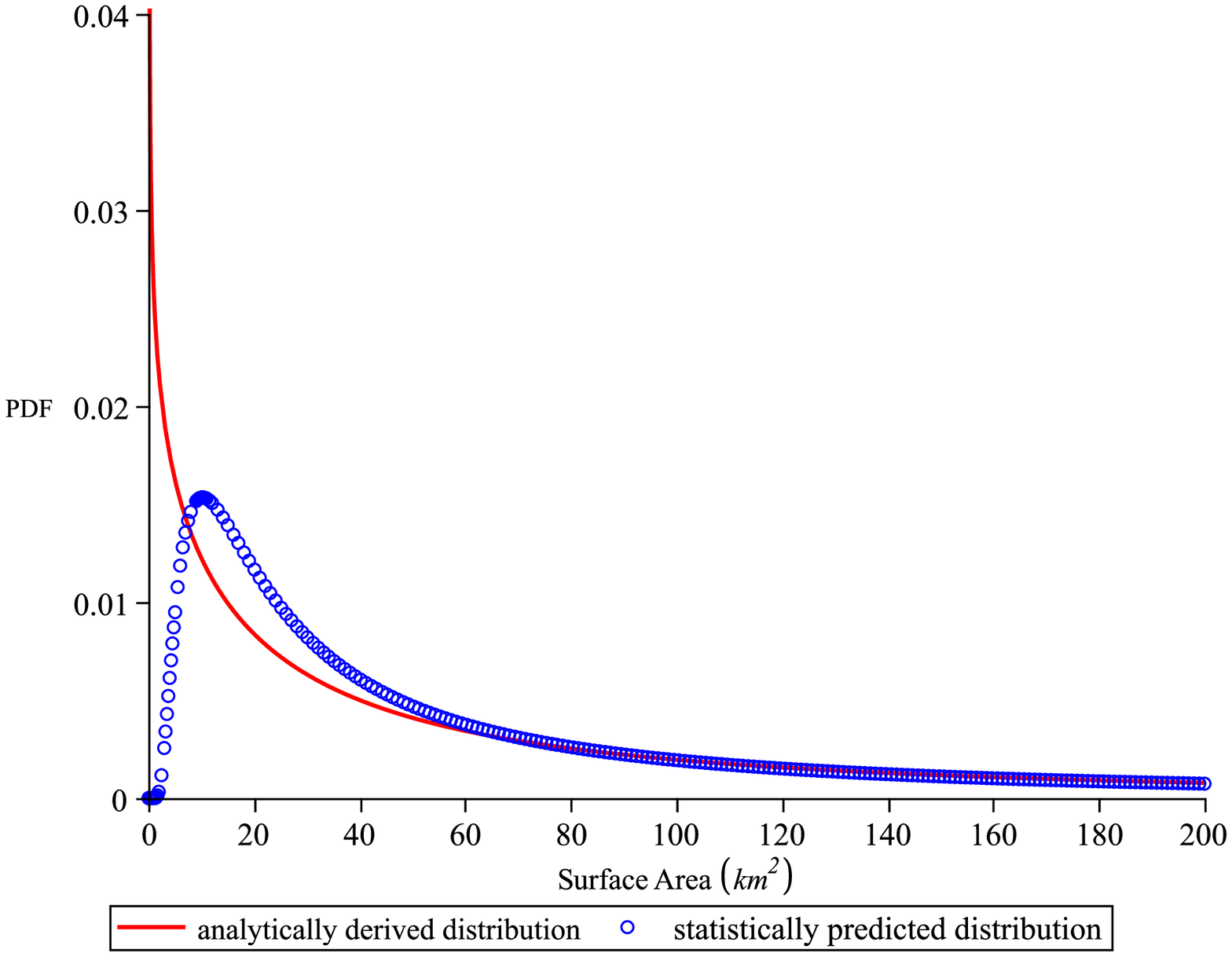}}
  \caption{(a) shows the statistically predicted and analytically derived PDF curves of $R$ based on $V$. (b) shows the statistically predicted and analytically derived PDF curves of $R$ based on $S$. (c) shows the statistically predicted and analytically derived PDF curves of $S$ based on $R$.
}
  \label{fig:subfig}
  \label{fig212}
\end{figure}

Because 5 irregular moons in the Pasiphae group were regrouped into the Ananke group, the distribution of the physical characteristics of this group underwent relatively large changes. As described in Section 3.3, the $p$-values corresponding to the physical characteristics of the log-logistic distribution are also very large, and the range of the confidence interval corresponding to the parameter values of the log-logistic distribution is relatively compact, so we believe that with the discovery of more moons belonging to this group, this distribution will be a potential distribution with a very high probability, not only because the current physical characteristics almost follow this distribution in the Ananke group and Carme group. Therefore, we now test the consistency of the physical characteristics of the Pasiphae group when they follow log-logistic distributions from the perspective of analytical derivation and statistical prediction.

According to Figures 13-16, it is found that the degree of agreement in the PDF curves between the statistically predicted and analytically derived distributions is much better than those shown in Figures 11 and 12, respectively.

%figure 13
   \begin{figure}[htb]
  \center
\begin{minipage}[htb]{165mm}
\vspace {2mm}
\centerline{\includegraphics[scale=0.55]{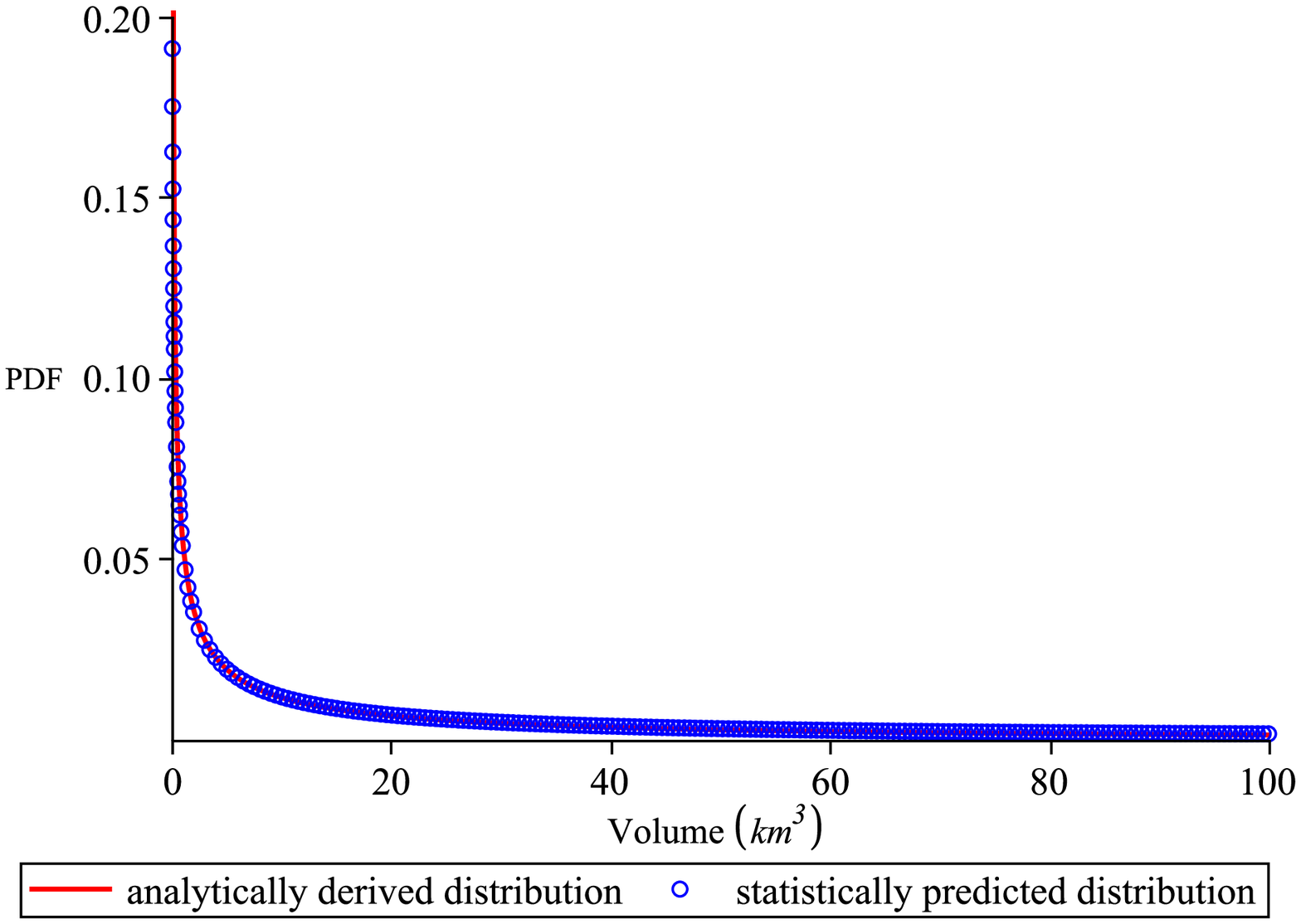}}
\vspace {-1mm}
\caption{Comparison of the PDF curves of $V$ between the statistically predicted and analytically derived distributions based on $R$.}
\label{fig215}
\end{minipage}
\end{figure}

 %figure 14
   \begin{figure}[htb]
  \center
\begin{minipage}[htb]{165mm}
\vspace {2mm}
\centerline{\includegraphics[scale=0.48]{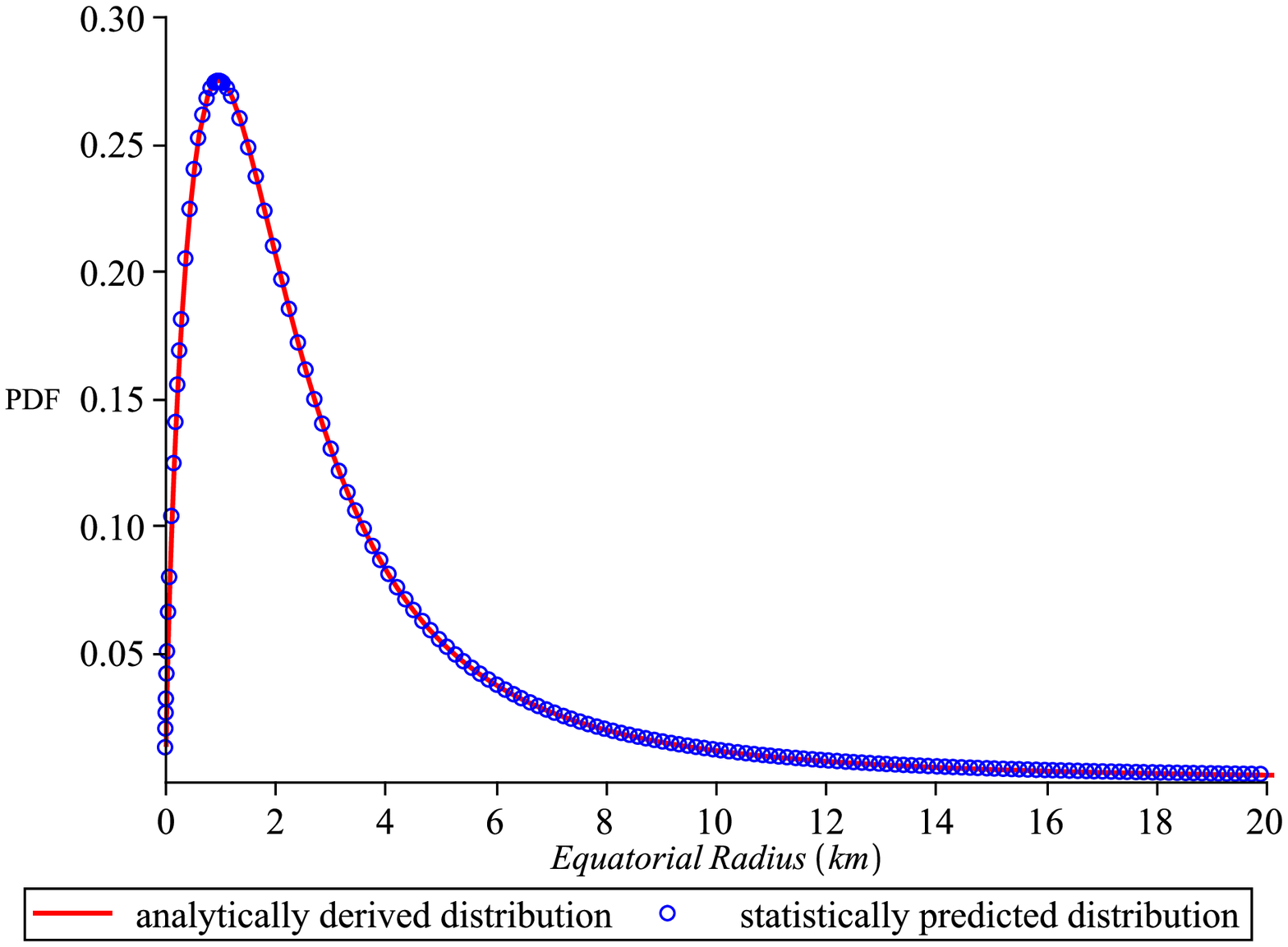}}
\vspace {-1mm}
\caption{Comparison of the PDF curves of $R$ between the statistically predicted and analytically derived distributions based on $V$.}
\label{fig216}
\end{minipage}
\end{figure}

 %figure 15
   \begin{figure}[htb]
  \center
\begin{minipage}[htb]{165mm}
\vspace {2mm}
\centerline{\includegraphics[scale=0.6]{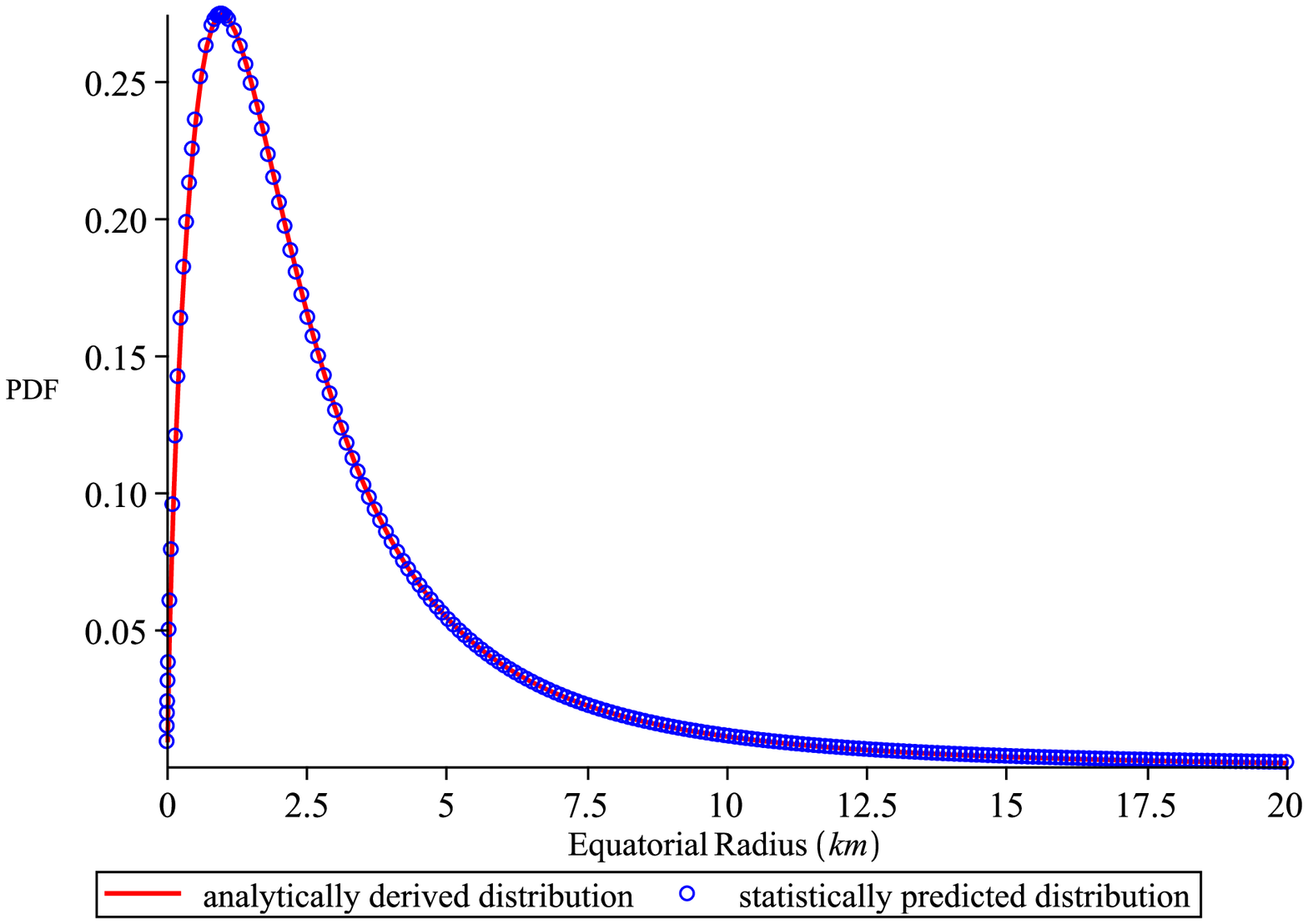}}
\vspace {-1mm}
\caption{Comparison of the PDF curves of $R$ between the statistically predicted and analytically derived distributions based on $S$.}
\label{fig217}
\end{minipage}
\end{figure}

 %figure 16
   \begin{figure}[htb]
  \center
\begin{minipage}[htb]{165mm}
\vspace {2mm}
\centerline{\includegraphics[scale=0.6]{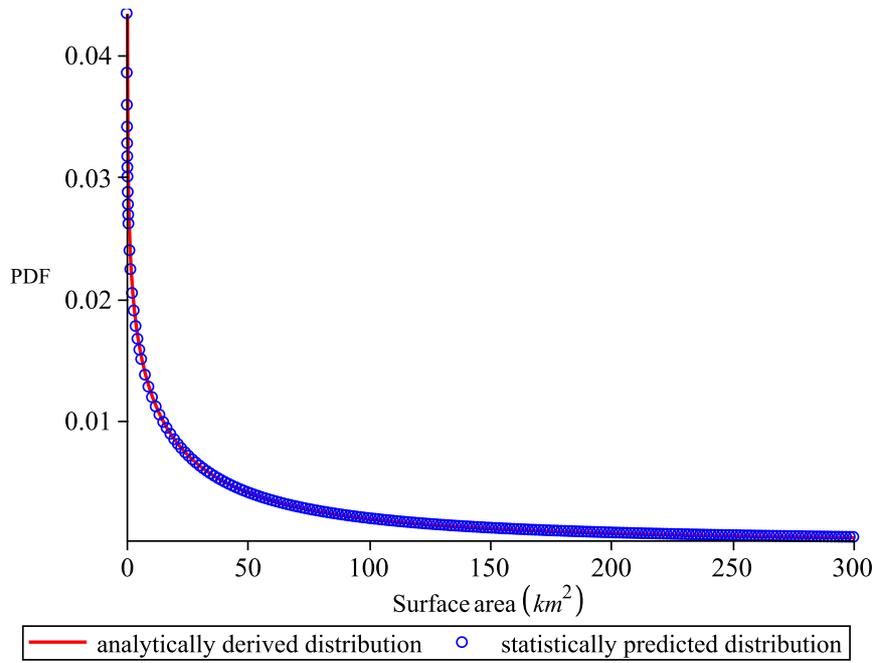}}
\vspace {-1mm}
\caption{Comparison of the PDF curves of $S$ between the statistically predicted and analytically derived distributions based on $R$.}
\label{fig218}
\end{minipage}
\end{figure}

\clearpage

\section{Conclusions}
Based on 21 commonly used continuous distributions, we apply K-S tests and maximum likelihood estimation to analyze the best-fit distributions of seven physical characteristics of the irregular moons in Jupiter's three major groups. These seven physical characteristics of the moons are the equatorial radius, equatorial circumference, volume, surface area, surface gravity, mass and escape velocity. The results of the statistical inference show that all seven physical characteristics of the moons in the Ananke group follow log-logistic distributions, and six physical characteristics of the moons in the Carme group also follow this distribution except the escape velocity, which follows a $t$ location-scale distribution or log-logistic distribution. In addition, more than half of the physical characteristics of the moons in the Pasiphae group follow log-logistic distributions. This phenomenon may be due to the clear decrease in the number of moons in the Pasiphae group after regrouping. We believe that this phenomenon will become more apparent as more Jupiter's moons belonging to the Pasiphae group are discovered. Compared with the results in [2], it is found that the distributions in this paper are much better and more consistent in describing the physical characteristics of Jupiter's moons, and the $p$-value is much larger in the current Pasiphae group.

Considering that all the physical features are not necessarily independent, the PDFs of relevant physical characteristics can also be obtained through strict analytical derivation and compared with the statistical prediction results. Therefore, the rationality of these distributions is further proved to some extent, especially the log-logistic distribution in this paper, which can describe the physical characteristics of most of Jupiter's irregular moons well, and we expect that this distribution will be helpful for future research on Jupiter's moons that are poorly understood or have not been discovered.

\appendix
\renewcommand\thetable{\Alph{section}\arabic{table}}
\section{Results of statistical inference}
See Tables A1-A12.
\clearpage
\setcounter{table}{0}
%Ananke: Table A1-Equatorial redius
\begin{landscape}
\begin{table}[]
\tiny
 \renewcommand{\arraystretch}{1.09}
\centering
\caption{Statistical inference results for the physical characteristics of the irregular moons in the Ananke group}
\scalebox{0.92}{
}
\end{table}
\end{landscape}

\clearpage
%Ananke: Table A2-Volume
\begin{landscape}
\begin{table}[]
 \renewcommand{\arraystretch}{1.09}
\tiny
\centering
\caption{Statistical inference results for the physical characteristics of the irregular moons in the Ananke group (Cont'd)}
\scalebox{0.92}{
%
}
\end{table}
\end{landscape}

\clearpage
%Ananke: Table A3-surface gravity
\begin{landscape}
\begin{table}[]
 \renewcommand{\arraystretch}{1.09}
\tiny
\centering
\caption{Statistical inference results for the physical characteristics of the irregular moons in the Ananke group (Cont'd) }
\scalebox{0.92}{
%
}
\end{table}
\end{landscape}

\clearpage
%Ananke: Table A4-Escape velocity
\begin{landscape}
\begin{table}[]
 \renewcommand{\arraystretch}{1.09}
\tiny
\centering
\caption{Statistical inference results for the physical characteristics of the irregular moons in the Ananke group (Cont'd)}
\scalebox{0.92}{
%
}
\end{table}
\end{landscape}

\clearpage
% Carme: Table A5-Equatorial radius
\begin{landscape}
\begin{table}[]
 \renewcommand{\arraystretch}{1.07}
\tiny
\centering
\caption{Statistical inference results for the physical characteristics of the irregular moons in the Carme group }
\scalebox{0.92}{
%
}
\end{table}
\end{landscape}

\clearpage
% Carme: Table A6-Volume
\begin{landscape}
\begin{table}[]
 \renewcommand{\arraystretch}{1.07}
\tiny
\centering
\caption{Statistical inference results for the physical characteristics of the irregular moons in the Carme group (Cont'd) }
\scalebox{0.92}{
%
}
\end{table}
\end{landscape}

\clearpage
% Carme: Table A7-Surface gravity
\begin{landscape}
\begin{table}[]
 \renewcommand{\arraystretch}{1.08}
\tiny
\centering
\caption{Statistical inference results for the physical characteristics of the irregular moons in the Carme group (Cont'd) }
\scalebox{0.9}{
%
}
\end{table}
\end{landscape}

\clearpage
% Carme: Table A8-Escape velocity
\begin{landscape}
\begin{table}[]
 \renewcommand{\arraystretch}{1.08}
\tiny
\centering
\caption{Statistical inference results for the physical characteristics of the irregular moons in the Carme group (Cont'd) }
\scalebox{0.92}{
%
}
\end{table}
\end{landscape}

\clearpage
% Pasiphae: Table A9-Equatorial radius
\begin{landscape}
\begin{table}[]
 \renewcommand{\arraystretch}{1.07}
\tiny
\centering
\caption{Statistical inference results for the physical characteristics of the irregular moons in the Pasiphae group}
\scalebox{0.92}{
%
}
\end{table}
\end{landscape}

\clearpage
% Pasiphae: Table A10-Volume
\begin{landscape}
\begin{table}[]
 \renewcommand{\arraystretch}{1.05}
\tiny
\centering
\caption{Statistical inference results for the physical characteristics of the irregular moons in the Pasiphae group (Cont'd)}
\scalebox{0.92}{
%
}
\end{table}
\end{landscape}

\clearpage
% Pasiphae: Table A11-Surface gravity
\begin{landscape}
\begin{table}[]
 \renewcommand{\arraystretch}{1.16}
\tiny
\centering
\caption{Statistical inference results for the physical characteristics of the irregular moons in the Pasiphae group (Cont'd)}
\scalebox{0.87}{
%
}
\end{table}
\end{landscape}

\clearpage
% Pasiphae: Table A12-Escape velocity
\begin{landscape}
\begin{table}[]
 \renewcommand{\arraystretch}{1.16}
\tiny
\centering
\caption{Statistical inference results for the physical characteristics of the irregular moons in the Pasiphae group (Cont'd)}
\scalebox{0.92}{
%
 \\
  \bottomrule
\end{tabular}}
\end{table}
\end{landscape}

\clearpage\noindent \textbf{Data Availability}\\
The data used to support the findings of this study are available from the corresponding author upon request.\\

 \noindent \textbf{Author Contributions}\\
Formal analysis, Fabao Gao; Software, Xia Liu; Writing---original draft, Xia Liu; Writing---review $\&$ editing, Fabao Gao.\\
 
\noindent  \textbf{Funding}\\
This research was funded by the National Natural Science Foundation of China (NSFC) though grant No.11672259 and the China Scholarship Council through grant No.201908320086.\\

\noindent  \textbf{Conflicts of Interest}\\
The authors declare that there is no competing interests.\\

\noindent\textbf{Acknowledgments}\\
We thank Dr. Abedin\,Y.\,Abedin whose comments and suggestions helped improve and clarify this manuscript.

\end{document}